%% file: Arxiv.tex
\documentclass[twocolumn,prb,notitlepage,floats,superscriptaddress,amsmath,amssymb]{revtex4-2}

\usepackage[version=3]{mhchem} 
\usepackage{bm}
\usepackage[utf8]{inputenc}
\usepackage[T1]{fontenc}
\usepackage{graphicx}
\usepackage{upgreek}
\usepackage{color}
\usepackage{ulem}
\usepackage{hyperref} 
\hypersetup{colorlinks,citecolor=blue, filecolor=blue ,linkcolor=blue , urlcolor=blue, pdftex}
\usepackage{sidecap}
\usepackage{amssymb}
\usepackage{braket}
\usepackage[caption=false]{subfig}

\def \FUW{Institute of Experimental Physics, Faculty of Physics, University of Warsaw, 02-093 Warsaw, Poland}
\def \Praga{Department of Condensed Matter Physics, Faculty of Mathematics and Physics, Charles University, CZ-121 16 Prague, Czech Republic}
\def \Wroclaw{Department of Semiconductor Materials Engineering, Wrocław University of Science and Technology, 50-370 Wrocław, Poland}
\def \Watanabe{Research Center for Functional Materials, National Institute for Materials Science, Tsukuba 305-0044, Japan}
\def \Taniguchi{International Center for Materials Nanoarchitectonics, National Institute for Materials Science, Tsukuba 305-0044, Japan}
 
\def \MagdaGk{Institute for Functional Intelligent Materials, National University of Singapore, 117544, Singapore}

\begin{document}

\title{Analogy and dissimilarity of excitons in monolayer and bilayer of MoSe$_2$}

\author{\L{}ucja Kipczak}
\email{lucja.kipczak@fuw.edu.pl}
\affiliation{\FUW}
\author{Artur O. Slobodeniuk}
\email{aslobodeniuk@karlov.mff.cuni.cz}
\affiliation{\Praga}
\author{Tomasz Wo\'zniak}
\affiliation{\Wroclaw}
\author{Mukul Bhatnagar}
\affiliation{\FUW}
\author{Natalia Zawadzka}
\affiliation{\FUW}
\author{Katarzyna Olkowska-Pucko}
\affiliation{\FUW}
\author{Magdalena Grzeszczyk}
\affiliation{\FUW}
\affiliation{\MagdaGk}
\author{Kenji~Watanabe}
\affiliation{\Watanabe}
\author{Takashi Taniguchi}
\affiliation{\Taniguchi}
\author{Adam Babi\'nski}
\affiliation{\FUW}
\author{Maciej R. Molas}
\email{maciej.molas@fuw.edu.pl}
\affiliation{\FUW}

\begin{abstract}
Excitons in thin layers of semiconducting transition metal dichalcogenides are highly subject to the strongly modified Coulomb electron-hole interaction in these materials.
Therefore, they do not follow the model system of a two-dimensional hydrogen atom. 
We investigate experimentally and theoretically excitonic properties in both the monolayer (ML) and the bilayer (BL) of MoSe$_2$ encapsulated in hexagonal BN.
The measured magnetic field evolutions of the reflectance contrast spectra of the MoSe$_2$ ML and BL allow us to determine $g$-factors of intralayer A and B excitons, as well as the $g$-factor of the interlayer exciton. 
We explain the dependence of $g$-factors on the number of layers and excitation state using first principles calculations.
Furthermore, we demonstrate that the experimentally measured ladder of excitonic $s$ states in the ML can be reproduced using the $\mathbf{k\cdot p}$ approach with the Rytova-Keldysh potential that describes the electron-hole interaction.
In contrast, the analogous calculation for the BL case requires taking into account the out-of-plane dielectric response of the MoSe$_2$ BL.
\end{abstract}

\maketitle

\section{Introduction \label{sec:Intro}}
The optical response of monolayers (MLs) belonging to semiconducting transition metal dichalcogenides (\mbox{S-TMDs}), such as MoS$_2$, MoSe$_2$, MoTe$_2$, WS$_2$, and WSe$_2$, is dominated by the excitonic emission/absorption even at room temperature~\cite{Mak2010, Arora2015, Arora2015b, molasNanoscale}. 
This is due to the binding energies (BE) of excitons, $i.e.$ bound electron-hole ($e$-$h$) pairs, which are as large as a few hundred meV ~\cite{Chernikov2014, Koperski2017, Wang2018, Molas2019Energy, Goryca2019}. 
The most unconventional property of these excitons is their non-Rydberg model spectrum, which cannot be described by the standard two-dimensional (2D) hydrogen atom~\cite{Macdonal1986, Koteles1988}.
A typical approach to account for the excitonic spectra of S-TMD MLs refers to the numerical solutions of the Schr\"{o}dinger equation, in which the $e$-$h$ attraction is approximated by the Rytova-Keldysh (RK) potential~\cite{Rytova1967, Keldysh1979}.
Although this is a well-known method for the ML, the corresponding analysis of the excitonic ladder for the bilayer (BL) 
has not been reported so far.

Moreover, the electronic bands in a BL and other 2H-stacked multilayers, 
are known to be effectively modified compared to the ML case~\cite{Fang2015, Sun2016, Ruiz2018, Molas2017Nanoscale, Slobodeniuk2019, Leisgang2020}.
This, in particular, implies the indirect bandgap in multilayers, which strongly affects their emission spectra~\cite{Mak2010, Arora2015, Arora2015b, molasNanoscale}.
Instead, more subtle effects of the hybridisation of electronic states around the direct bandgap in multilayers are relevant for absorption-type processes~\cite{Horng2018, Slobodeniuk2019, Paradisanos2020}.
Recently, it has been shown that the absorption resonances due to the interlayer/hybridised excitons observed in the bi- and trilayers of S-TMDs can be widely tuned using the external electric field applied perpendicularly to the layers plane~\cite{Leisgang2020, Das2020, Lorchat2021, Zhao2022, Altaiary2022, Hagel2023, Shun2022}, which understanding may be also of importance in potential applications.

The external out-of-plane magnetic field is widely applied to study the thin layers of S-TMDs due to their significant Zeeman response~\cite{Koperski2019, Arora2018}. 
Consequently, the measured emission/absorption resonances split into two circularly polarised components with the magnitude of the Zeeman splitting denoted by the $g$-factor.
Because of the $g$-factor values found in S-TMD MLs, different types of transition can be identified: bright ($g$-factor about -4), spin- (-8) and momentum-forbidden dark (-12)~\cite{Liu2020, He2020, Zinkiewicz2021, Pucko2023}.
In the case of BL, the found $g$-factors are more scattered.
As for intralayer excitons in S-TMD BLs, the found $g$-factors are also about -4~\cite{Li2014, Koperski2019, Arora2018}, the corresponding $g$-factors for interlayer excitons are around 8~\cite{Slobodeniuk2019}.
Then, the \mbox{$g$-factor} not only gives us information concerning the Zeeman effect, but can also be a method to identify different excitonic complexes in thin layers of S-TMDs.
\begin{figure*}[t]
		\subfloat{}%
		\centering
		\includegraphics[width=1 \linewidth]{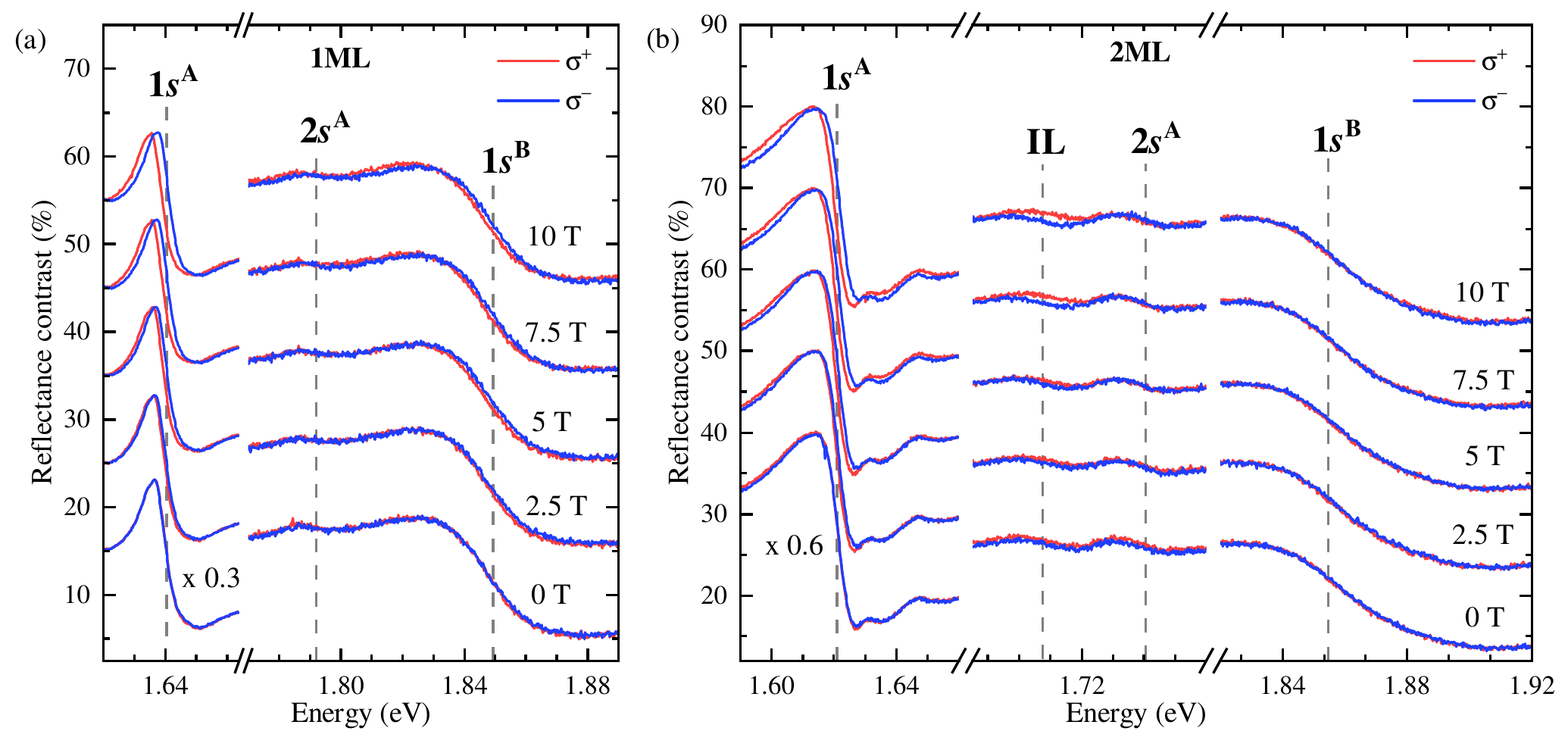}
    	\caption{Helicity-resolved RC spectra of (a) monolayer and (b) bilayer of MoSe$_2$ encapsulated in hBN for selected values of magnetic field measured at $T$=10~K. 
        The red and blue curves correspond to $\sigma^+$ and $\sigma^-$ polarizations of reflected light in magnetic fields applied perpendicularly to the layers plane, respectively. 
        The spectra are vertically shifted for clarity.}
		\label{fig:rc}
\end{figure*}

In this work, we investigate experimentally and theoretically excitonic properties in high-quality ML and BL of molybdenum diselenide (MoSe$_2$) encapsulated in hexagonal boron nitride (hBN).
The low-temperature ($T$=10~K) reflectance contrast (RC) spectra are measured in external magnetic fields up to 10 T, applied in the out-of-plane configuration.
Resonances, related to both neutral A and B excitons, are identified in ML and BL.
Moreover, the transition associated with the interlayer exciton (IL) was recognised in the RC spectra of BL.
The excitation ladder of A excitons in the ML and BL limits is modelled based on the $\mathbf{k\cdot p}$ approach using the modified RK potential. 
Furthermore, the experimentally obtained Land\'e $g$-factors of the excitonic resonances are explained by first principles calculations.
Apart from the Introduction and Summary sections, our paper is composed of three (2–4) main sections completed by the Methods section and the Supplemental Material (SM).
First, in Sec.~\ref{sec:results}, we focus on the analysis of the magnetic field evolutions of the reflectance contrast spectra measured on the MoSe$_2$ ML and BL. 
The theoretical approach for excitation spectra of excitons in the S-TMD ML and BL using the $\mathbf{k\cdot p}$ method is presented in Sec.~\ref{sec:excitons}.
Section~\ref{sec:gfactors} is dedicated to calculations of the excitonic $g$-factors in the MoSe$_2$ ML and BL.

\section{Experimental results\label{sec:results}}

Figs.~\ref{fig:rc}(a) and (b) present the RC spectra measured on the ML and BL of MoSe$_2$ encapsulated in hBN at the selected values of applied \mbox{out-of-plane magnetic fields}. 
The two main resonances, labeled correspondingly 1$s^\textrm{A}$ and 1$s^\textrm{B}$ in the figure, are associated with the absorption processes of the ground $s$ states of the \mbox{intralayer} A and B excitons~\cite{aroramose2, Koperski2017}.
The appearance of the A and B excitons is due to a relatively large spin-orbit splitting in the MoSe$_2$ VB~\cite{Koperski2017}.
For the ML, the first excited $s$ state of the A exciton (2$s^\textrm{A}$) is also observed with a substantially lower intensity compared to the main 1$s^\textrm{A}$ and 1$s^\textrm{B}$ resonances.
There are two less pronounced transitions apparent in the RC spectra measured on the BL, denoted 2$s^\textrm{A}$ and IL, which we attribute correspondingly to the first excited state of the A exciton and to the interlayer exciton. 
The identification of the 1$s^\textrm{A}$, 1$s^\textrm{B}$, and 2$s^\textrm{A}$ resonances, observed in the RC spectrum of the MoSe$_2$ ML, is straightforward and is in accordance with many other studies on MoSe$_2$ MLs encapsulated in hBN~\cite{Molas2019Energy, Goryca2019, Xiao2021}.
Note that the measured low-temperature RC spectrum of the MoSe$_2$ BL is similar to that reported in Ref.~\cite{Horng2018}.

As can be appreciated from Fig.~\ref{fig:rc}, all the observed resonances split into two circularly polarised components under the applied out-of-plane magnetic field due to the Zeeman effect~\cite{Koperski2019}.
The intralayer A- and B-related resonances are characterised by the same sign of the splitting ($\sigma^+$ energy is lower than the $\sigma^-$ one), but with different magnitude.
At the same time, the splitting of the IL transition is opposite ($\sigma^-$ energy is lower than $\sigma^+$ one).
In order to investigate in detail the magnetic field evolution of the observed resonances, we fitted them using the Fano-type function.
Note that, as we have not performed the analysis within the framework of the transfer matrix method combined with the Lorentz oscillator model~\cite{molasNanoscale}, the extracted energy evolutions as a function of magnetic field are biased, particularly, for the resonances with small intensity (2$s^\textrm{A}$, IL). 
A detailed description of the A, B and IL excitons, and their optical selection rules in terms of the electronic band structure can be found in the SM.

\begin{figure}[t]
		\subfloat{}%
		\centering
		\includegraphics[width=1 \linewidth]{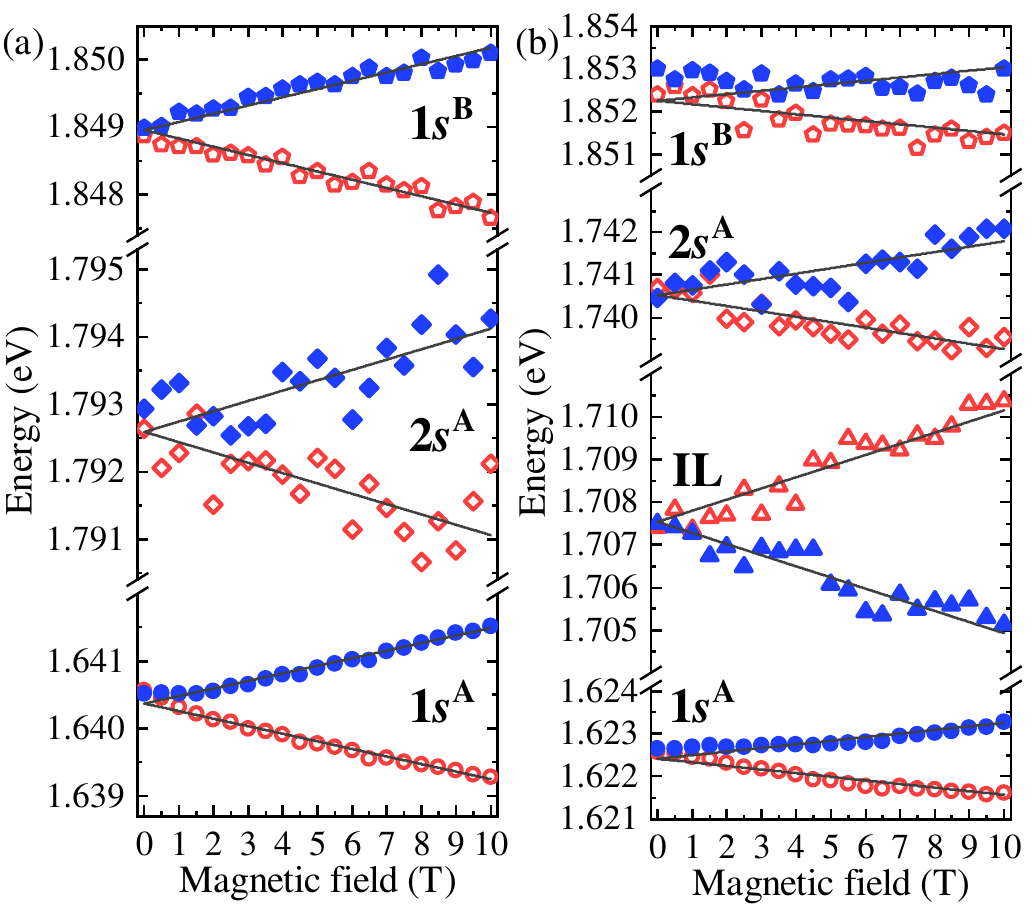}
    	\caption{Obtained excitonic energies of the $\sigma^\pm$ components of the transitions measured on (a) monolayer and (b) bilayer of MoSe$_2$  encapsulated in hBN as a function of the out-of-plane magnetic field.
         The red and blue points correspond to $\sigma^+$ and $\sigma^-$ polarizations, respectively.
         The solid black lines represent fits according to the equation described in the text.}
		\label{fig:zeeman}
\end{figure}

The field evolution for the $\sigma^\pm$ components of the investigated transitions is presented in Fig.~\ref{fig:zeeman}.
Upon application of an out-of-plane magnetic field ($B$), the magnetic-field dependences of the $\sigma^\pm$ energies ($E_{\sigma^\pm}$) can be defined as $E_{\sigma^\pm}(B)=E_0 \pm 1/2 g \mu_\textrm{B} B$, where $E_0$ is the transition energy at zero field, $g$ denotes the $g$-factor of the considered resonance, $\mu_\textrm{B}$ is the Bohr magneton.
The fitted curves are presented in Fig.~\ref{fig:zeeman}.
The $E_0$ energies and the $g$-factors for all the studied transitions are summarised in Tab.~\ref{tab:g-factors}.
To verify the obtained values of the $g$-factors, we also calculate theoretically the corresponding $g$-factors using the Density Functional Theory (DFT) method, see Sec.~\ref{sec:gfactors} for details.
The experimentally determined $g$-factors for the ground states of the A (B) excitons, $i.e.$ 1$s^\textrm{A}$ (1$s^\textrm{B}$), in the ML and BL of MoSe$_2$ equal to $-3.87 \pm 0.01$ ($-4.22 \pm 0.11$) and $-2.92 \pm 0.02$ ($-2.71 \pm 0.17$), respectively.
These values agree very well with those theoretically calculated (see Tab.~\ref{tab:g-factors}) and those previously reported for the intralayer A and B excitons in the MoSe$_2$ ML and BL~\cite{Li2014, Koperski2019, Arora2018, Goryca2019}.
The $g$-factors found for the first excited states (2$s^\textrm{A}$) of the A exciton in both ML and BL are $-5.28 \pm 0.41$ and $-4.36 \pm 0.24$, respectively.
It is interesting that the magnitudes of the 2$s$ $g$-factors are significantly larger (of about 40$\%$-50$\%$) than those of the 1$s^\textrm{A}$ states.
In contrast, the theoretically calculated values of the $g$-factors for the 1$s$ and 2$s$ states of the A/B excitons are larger/smaller by about a few percent.
Note that both possible changes ($i.e.$ increase or decrease) of the corresponding $g$-factors of the $s$ states are reported in the literature~\cite{Stier2016, Chen2019, Delhomme2019, Molas2019Energy, Goryca2019, Raiber2022}.
In our opinion, the discrepancy between the experimental and theoretical values of the 2$s^\textrm{A}$ $g$ factor requires a more sophisticated analysis, which goes beyond the scope of this work.
The value of $g$-factor for the IL transition of $9.01 \pm 0.34$, which has not been reported so far, is in very good agreement with the theoretically calculated values of 8.71 and 8.91 for the IL$^\textrm{A}$ and IL$^\textrm{B}$ excitons, respectively. 
As the IL$^\textrm{A}$ resonance is reported mainly in MoS$_2$ and MoSe$_2$ BLs~\cite{Horng2018, Slobodeniuk2019, Gerber2019, Paradisanos2020, Leisgang2020, Grzeszczyk2021, Shun2022}, we ascribe the observed IL resonance to the IL$^\textrm{A}$.
Further discussion on the interlayer excitons can be found in the SM.

\begin{table}[t]
\caption{\label{tab:g-factors}
Experimentally determined values of the $E_0$ energies and the $g$-factors for all the studied transitions. The $g^{calc}$ values correspond to the theoretically calculated parameters using DFT method.
Notably, the experimental values of the $E_0$ energy and the $g$-factor are assumed to be obtained for the IL$^\textrm{A}$ exciton.}
\begin{tabular}{cccc}
      & $E_0$ & $g$  & $g^{calc}$\\
\hline
\multicolumn{4}{c}{monolayer} \\
\hline
1$s^\textrm{A}$  & $1.641 \pm 0.001$  & $-3.87 \pm 0.01$ &  $-3.69$ \\
2$s^\textrm{A}$  & $1.793 \pm 0.005$  & $-5.28 \pm 0.41$ &  $-3.90$ \\
1$s^\textrm{B}$  & $1.849 \pm 0.003$  & $-4.22 \pm 0.11$ &  $-3.75$ \\
\hline
\multicolumn{4}{c}{bilayer} \\
\hline
1$s^\textrm{A}$   & $1.623 \pm 0.001$  & $-2.92 \pm 0.02$ &  $-3.00$ \\
2$s^\textrm{A}$   & $1.741 \pm 0.006$  & $-4.36 \pm 0.24$ &  $-3.19$ \\
1$s^\textrm{B}$   & $1.853 \pm 0.003$  & $-2.71 \pm 0.17$ &  $-3.16$ \\
IL$^\textrm{A/B}$ & $1.707 \pm 0.006$  & $+9.01 \pm 0.34$ &  $+8.71/+8.91$ \\
\end{tabular}
\end{table}

\section{Excitonic ladder in monolayer and bilayer: theoretical approach\label{sec:excitons}}
A common approach to account for the excitation spectra of excitons in S-TMD MLs refers to the numerical solutions of the Schr{\"o}dinger equation, in which the \mbox{$e$-$h$} attraction is approximated by RK potential~\cite{Rytova1967, Keldysh1979}. 
Although this numerical method gives very good results for Mo- and W-based MLs~\cite{Cudazzo2011, Stier2016, Goryca2019}, there is a lack of analogous calculations of the excitation spectra of excitons in multilayer systems (particularly, in a BL).
In the following, the calculations of the excitonic ladders of the $s$ states in both the ML and the BL of the \mbox{S-TMDs} are presented.
In order to verify our theoretical results, we compare them with the experimentally found energy separation between the 1$s$ and 2$s$ states, $\Delta E^{exp}_{12}$, obtained using the values shown in Tab.~\ref{tab:g-factors}.
Note that our comprehensive approach to the theoretical analysis of the excitation spectra of excitons in S-TMD ML and BL is described in detail in the SM.

First, we calculate a spectrum of the intralayer excitons in the ML and BL with the help of the effective \mbox{two-body} hydrogen-like problem. 
We derive such a problem from the band Hamiltonians of the considered 2D systems within the $\mathbf{k\cdot p}$ approximation.  
To do this, we first consider the basic electronic properties of S-TMD ML.
The ML crystal is a direct bandgap semiconductor. 
The extrema of the valence (VB) and conduction (CB) bands are located at the $\mathrm{K}^\pm$ points of the Brillouin zone (BZ).
Due to the strong spin-orbit interaction, both bands are spin-split. 
The values of the splittings in the VB ($\Delta_v$) and in the CB ($\Delta_c$) are hundreds and tens of meV, respectively~\cite{Kormanyos2015}.
Therefore, the Bloch states at the $\mathrm{K}$ points can be presented as a tensor product of the spin $\{|\downarrow\rangle, |\uparrow\rangle\}$ and the spinless band states.  
The spinless VB and CB states at $\mathrm{K}^\pm$ points are made predominantly from $d_{x^2-y^2}\pm id_{xy}$ and $d_{z^2}$ orbitals of transition metal atoms, respectively~\cite{Gong2013, Kormanyos2015}. 
Such a structure of Bloch states in opposite K points defines the optical selection rules in the ML and is a consequence of the system's time-reversal symmetry (TRS).
The TRS also dictates that Bloch states with the same band index ($c$ or $v$) but with opposite spins in opposite valleys have equal dispersion laws, $i.e.$ the same band structure. 
Therefore, we restrict our consideration of the conduction and valence bands to the K$^+$ point for brevity. 
All conclusions for the K$^-$ point can be done by analogy. 

The A excitons at the K$^+$ point of the ML are formed from an electron from the bottom spin-up CB and \mbox{a hole} from the top spin-up VB. 
The corresponding spinless Bloch functions are $|\Psi_c\rangle$ and $|\Psi_v\rangle$. 
The two-band fully-diagonalized Hamiltonian of these bands, written in the corresponding basis
$\{|\Psi_c\rangle\otimes|\uparrow\rangle, |\Psi_v\rangle\otimes|\uparrow\rangle\}$, can be presented in the form 
\cite{Slobodeniuk2022}
\begin{equation}
\label{eq:HamiltonianML}
H_{ML}=\left[
\begin{array}{cc}
E_g +\frac{\hbar^2\mathbf{k}^2}{2m_e} & 0  \\
0 & -\frac{\hbar^2\mathbf{k}^2}{2m_h} \\
\end{array}
\right]. 
\end{equation}
Here, $E_g$ is the bandgap energy parameter, 
$\mathbf{k}=k_x\mathbf{e}_x+k_y\mathbf{e}_y$ is the in-plane momentum of the quasiparticles in the ML, where $\mathbf{e}_x,\mathbf{e}_y$ are unit vectors in the $x$ and $y$ directions, respectively.
$m_e, m_h>0$ are correspondingly the electron and hole effective masses in the ML. 
The Rydberg-type spectrum of $e$-$h$ pairs for such a band structure can be found from the solution of the corresponding eigenvalues problem \cite{Wannier1937,Keldysh1979}
\begin{equation}
\label{eq:eigenvalues_monolayer}
\Big\{-\frac{\hbar^2}{2\mu}\nabla_\parallel^2+V_{RK}(\rho)\Big\}\psi(\rho)=E\psi(\rho),
\end{equation}
where $\mu=m_em_h/(m_e+m_h)$ is the reduced exciton mass, 
$\nabla_\parallel=\mathbf{e}_x\partial_x+\mathbf{e}_y\partial_y$ is the 2D nabla operator, $\rho$ is the \mbox{in-plane} distance between an electron and a hole of the exciton, $E$ is the exciton energy, and $V_{RK}(\rho)$ is the Rytova-Keldysh potential \cite{Rytova1967,Keldysh1979}
\begin{equation}
\label{eq:Rytova_Keldysh}
V_{RK}(\rho)=-\frac{\pi e^2}{2r_0}\Big[\text{H}_0\Big(\frac{\rho\varepsilon}{r_0}\Big)-
Y_0\Big(\frac{\rho\varepsilon}{r_0}\Big)\Big].
\end{equation}
Here $\text{H}_0(x)$ and $Y_0(x)$ correspond to the Struve and Bessel functions of the second kind, $r_0=2\pi\chi_{2D}$ represents the screening length, $\chi_{2D}$ is the ML 2D polarizability \cite{Cudazzo2011, Berkelbach2013}. 
$\varepsilon$ denotes the dielectric constant of the surrounding medium (hBN in our study).

Let us consider the band structure of the BL at the K$^+$ point. 
To do this, we first define the Bloch states of the VB and CB at the $\mathrm{K}^+$ point of the BL, by constructing them from the ML Bloch states of the top and bottom layers of the BL.
Namely, we introduce the states $|\Psi^{(m)}_n\rangle\otimes|s\rangle$, where $m=1,2$ is a layer index
(for bottom and top layers, respectively), $n=v,c$ is a band index (for VB and CB) and $s=\uparrow,\downarrow$ specifies the spin degree of freedom. 
The bottom (first) layer states $|\Psi^{(1)}_v\rangle$ and $|\Psi^{(1)}_c\rangle$
are made predominantly from $d_{x^2-y^2}+id_{xy}$ and $d_{z^2}$ orbitals of transition metal atoms,
respectively~\cite{Gong2013,Kormanyos2015}. 
They coincide with ML spinless states $|\Psi_v\rangle$ and $|\Psi_c\rangle$, considered above. 
The top (second) layer states $|\Psi^{(2)}_v\rangle$ and $|\Psi^{(2)}_c\rangle$ are made from
$d_{x^2-y^2}- id_{xy}$ and $d_{z^2}$ orbitals and coincide with spinless states in the
$\mathrm{K}^-$ point of the ML. 
In our study, we suppose the orthogonality of the basis states from different layers and bands $\langle\Psi_n^{(m)}|\Psi_{n'}^{(m')}\rangle=\delta_{nn'}\delta_{mm'}$. 

The symmetry analysis of the BL system~\cite{Arora2018, Slobodeniuk2019, Grzeszczyk2021} demonstrates that the electron excitations of the CB of the different layers do not 
interact with each other in the leading order, $i.e.$ the electron states of the BL are localised either in the bottom or in the top layer. 
On the contrary, the VB of the different layers of the BL interact with each other forming the new VB with the Bloch states delocalised in the \mbox{out-of-plane} direction. 
These states can be found by diagonalizing the VB part of the BL Hamiltonian, written on the basis   
$\{|\Psi^{(1)}_v\rangle\otimes|\uparrow\rangle,|\Psi^{(2)}_v\rangle\otimes|\uparrow\rangle, 
|\Psi^{(2)}_v\rangle\otimes|\downarrow\rangle,|\Psi^{(1)}_v\rangle\otimes|\downarrow\rangle\}$ 
\begin{equation}
H^{VB}_{BL}=\left[
\begin{array}{cccc}
-\frac{\hbar^2\mathbf{k}^2}{2m_h} & t & 0 & 0 \\
t & -\Delta_v-\frac{\hbar^2\mathbf{k}^2}{2m_h} & 0 & 0  \\
0 & 0 & -\frac{\hbar^2\mathbf{k}^2}{2m_h} & t \\
0 & 0 &  t & -\Delta_v-\frac{\hbar^2\mathbf{k}^2}{2m_h}
\end{array}
\right], 
\end{equation}
where $t$ is the interlayer hopping term.
The spectrum of this Hamiltonian is doubly degenerated by spin (in full accordance with the TRS and inverse symmetry of the BL)
\begin{equation}
E_{VB}^\pm=-\frac{\hbar^2\mathbf{k}^2}{2m_h}-\frac{\Delta_v}{2}\pm \sqrt{\frac{\Delta_v^2}{4}+t^2}.
\end{equation}
The eigenstates that correspond to the upper-energy $E_{VB}^+$ bands are
\begin{align}
&|\Phi^+_{v\uparrow}\rangle=\Big[\cos\theta|\Psi^{(1)}_v\rangle + \sin\theta|\Psi^{(2)}_v\rangle\Big]\otimes|\uparrow\rangle, \\
&|\Phi^+_{v\downarrow}\rangle=\Big[\sin\theta|\Psi^{(1)}_v\rangle+
\cos\theta|\Psi^{(2)}_v\rangle\Big]\otimes|\downarrow\rangle,
\end{align}
where we introduced $\cos(2\theta)=\Delta_v/\sqrt{\Delta_v^2+4t^2}$.
Note that the new Bloch states describe the delocalised in the out-of-plane direction VB excitations. 
For example, the state $|\Phi^+_{v\uparrow}\rangle$ describes the VB excitation, which can be found with probabilities $P^{(1)}=\cos^2\theta$ and $P^{(2)}=\sin^2\theta$ in the first (bottom) and second (top) layers, respectively.
The eigenstates $|\Phi^-_{v\uparrow}\rangle$ and $|\Phi^-_{v\downarrow}\rangle$, which correspond to the lower-energy $E_{VB}^-$ bands, can be derived from the first ones by replacing $\cos\theta\rightarrow -\sin\theta, \sin\theta\rightarrow \cos\theta$.

The optical transitions in the K$^+$ point of the BL, which form the intralayer A-excitons in the BL, couple either
$\{|\Psi^{(1)}_c\rangle\otimes|\uparrow\rangle,|\Phi^+_{v\uparrow}\rangle\}$ or $\{|\Psi^{(2)}_c\rangle\otimes|\downarrow\rangle,|\Phi^+_{v\downarrow}\rangle\}$ group of the bands.
The transitions between the first pair of the bands are active in $\sigma^+$ polarised light, while the transitions between the second pair of the bands are active in $\sigma^-$ polarised light. 
The first and second groups of the bands are described by the same Hamiltonian 
\begin{equation}
H_{BL}=\left[
\begin{array}{cc}
E_g +\frac{\hbar^2\mathbf{k}^2}{2m_e} & 0  \\
0 & -\frac{\hbar^2\mathbf{k}^2}{2m_h}-\frac{\Delta_v}{2}+\sqrt{\frac{\Delta_v^2}{4}+t^2} \\
\end{array}
\right], 
\end{equation}
which looks similar to the two-band Hamiltonian of the ML, compare with Eq.~\ref{eq:HamiltonianML}. 
Note that the effective electron and hole masses of these bands coincide with the corresponding masses in the ML (in the leading order of the $\mathbf{k\cdot p}$ approximation). 
Therefore, the intralayer A-excitons are characterised by the same reduced mass $\mu$, as in the ML.
On the other hand, the delocalisation of the VB Bloch state in the out-of-plane direction leads to the modification of the Coulomb interaction between such a hole excitation and an electron excitation, which remains localised in one of the layers. 
The modified Coulomb interaction in the BL $V_{bil}(\rho)$ is derived in the SM. 
In summary, we conclude that the spectrum of the intralayer A-excitons can be derived from Eq.~(\ref{eq:eigenvalues_monolayer}) by replacing $V_{RK}(\rho)$ with $V_{bil}(\rho)$.

\subsection{Monolayer}
\begin{figure}[t]
		\subfloat{}%
		\centering
		\includegraphics[width=1 \linewidth]{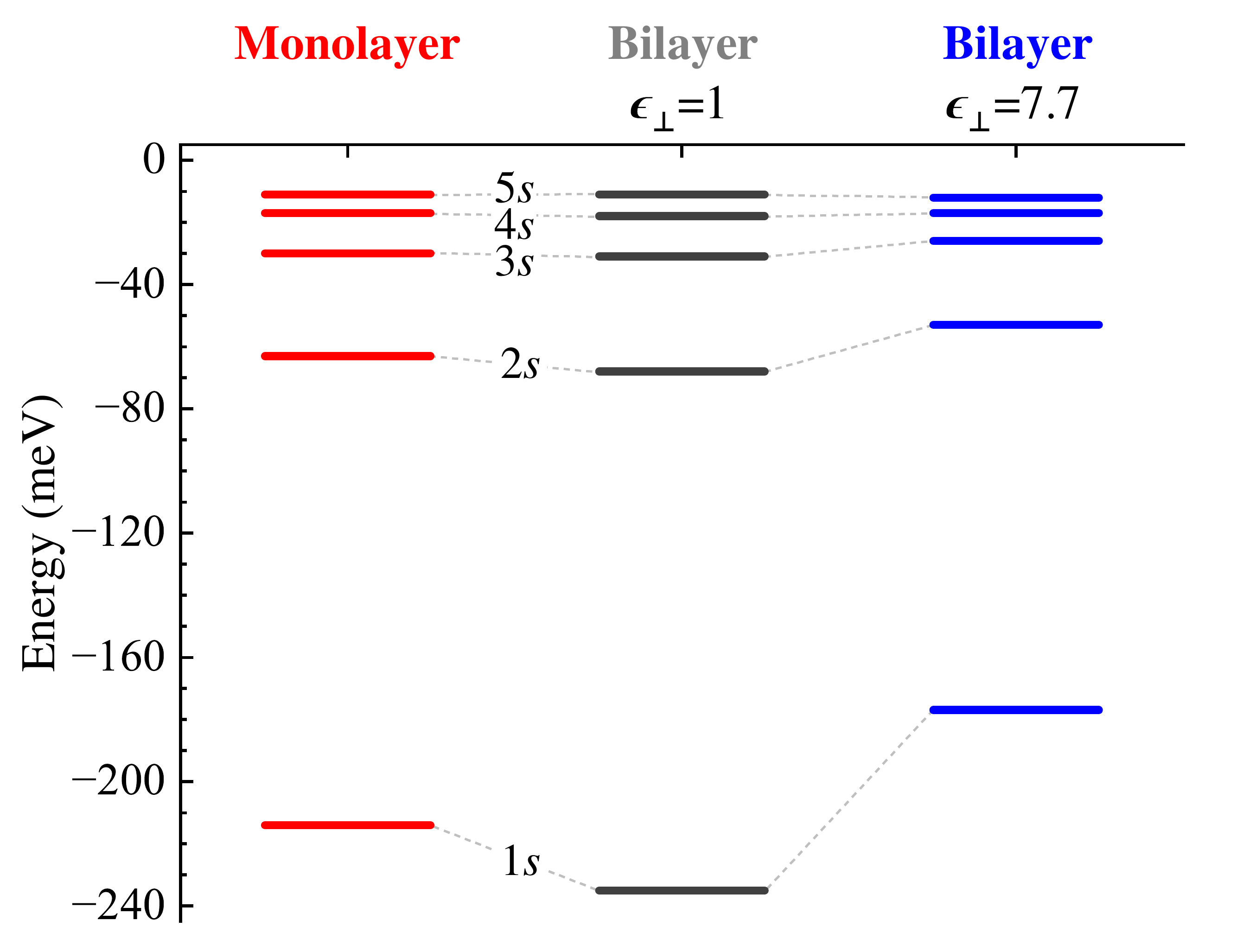}
    	\caption{Energy spectrum of $s$ excitonic state of the MoSe$_2$ ML and BL encapsulated in hBN calculated using the $\mathbf{k\cdot p}$ approach. The $\epsilon_\perp$ parameter denotes the out-of-plane dielectric constant used in calculations.}
		\label{fig:spectrum}
\end{figure}
We derive the spectrum of excitons in the ML by solving Eq.~(\ref{eq:eigenvalues_monolayer}) with Rytova-Keldysh potential (\ref{eq:Rytova_Keldysh}). 
Following Ref.~\cite{Molas2019_1}, we introduce the dimensionless parameters $\xi=\rho\varepsilon/r_0$ and $\epsilon=E/Ry^*$, with $Ry^*=\mu e^4/2\hbar^2\varepsilon^2$ and rewrite Eq.~(\ref{eq:eigenvalues_monolayer}) for the case of $s$-type excitons in the following form
\begin{equation}
\Big\{b^2\frac{1}{\xi}\frac{d}{d\xi}\Big(\xi\frac{d}{d\xi}\Big)+b v_{RK}(\xi)+\epsilon\Big\}\psi(\xi)=0,
\end{equation} 
where $b=\hbar^2\varepsilon^2/\mu e^2r_0$ and $v_{RK}(\xi)=\pi[\text{H}_0(\xi)-Y_0(\xi)]$. 
Note that the parameter $Ry^*$ defines the natural energy scale for the considered excitonic problem, see Ref.~\cite{Molas2019_1} for details. 
The dimensionless parameter $b$ is the ratio of the reduced Bohr radius \mbox{$a_0^*=\hbar^2\varepsilon/\mu e^2$} and the reduced screening length $r_0^*=r_0/\varepsilon$ in the system.
We use $\varepsilon=4.5$~\cite{geick1966}, $r_0=51.7\mbox{\AA}$, $\mu=0.44m_0$ for our calculations, yielding $b\approx 0.47$, see details in the SM.
The numerical solution of the equation for this value of $b$ gives the energies of the ground (1$s$) and the first four excited (2$s$, 3$s$, 4$s$, and 5$s$) states equal to $E_1=-214$\,meV, $E_2=-63$\,meV, $E_3=-30$\,meV, $E_4=-17$\,meV, \mbox{$E_5=-11$\,meV}, respectively.
The calculated energy positions of the excitonic states in ML are presented in Fig.~\ref{fig:spectrum}.
Note that the absolute energy of a given $ns$ excitonic state, $E_{n,ex}=E_\text{g}+E_n$, is difficult to predict since the bandgap energy ($E_\text{g}$) is renormalised by the Coulomb interaction and also depends on the dielectric constant $\varepsilon$. 
Calculating such a bandgap shift requires an additional more resource-demanding numerical investigation. 
To verify our theoretical calculations, we determine the energy distance between 1$s$ and 2$s$ emission lines $\Delta E_{12}=E_{1, ex}-E_{2, ex}=(E_\text{g}+E_1)-(E_\text{g}+E_2)$. 
Therefore, we found \mbox{$|\Delta E_{12}|=151$\,meV}, which is nearly perfectly consistent with our experimental value $|\Delta E_{12}^{exp} |=152\pm5$\,meV as well as the previously reported value of about 153~meV obtained from photoluminescence experiment~\cite{Molas2019Energy}.

\subsection{Bilayer}
The computation of the excitonic spectrum in the BL of S-TMDs is a much more sophisticated task. 
For the ML case, the charges and wavefunctions of an electron and a hole are confined within the ML plane.
The situation with the electron and hole electronic excitations in the BL is more complex. 
The hybridisation of the VB states leads to the charge redistribution of hole quasiparticles between layers in the BL~\cite{molasNanoscale, Slobodeniuk2019, Grzeszczyk2021}.
Using the properties of the VB states in the BL, we obtain the following values for the charges of the hole excitation, which belong to the same ($Q_{in}$) and opposite ($Q_{opp}$) layers as an electron, 
\begin{equation}
Q_{in/opp}=\frac{|e|}{2}\Big(1\pm\frac{\Delta_v}{\sqrt{\Delta_v^2+4t^2}}\Big)\approx 0.932|e|/0.068|e|,
\end{equation}
where "+" and "-" signs correspond to the same $(in)$ and opposite $(opp)$ layers of the BL, respectively.  
Here we used the numbers $\Delta_v=182$\,meV and $t=53$\,meV~\cite{Gong2013}.
Subsequently, the redistribution of the hole charge in the out of-plane-direction modifies the Coulomb potential between electron and hole excitations in the BL.
The derivation of the corresponding potential $V_{bil}(\rho)$ as a function of the in-plane distance, $\rho$ between an electron and hole excitation is presented in the SM. 
Note that the potential $V_{bil}(\rho)$ depends on the distance $L$ between the layers in the BL, the screening length $r_0$, the dielectric constant of the medium surrounding the bilayer $\varepsilon$, and finally the out-of-plane dielectric constant $\epsilon_\perp$ of the BL. 
The case $\epsilon_\perp=1$ corresponds to the limit situation when the BL can not be polarised by an electric field in an out-of-plane direction.
In the real situation the BL, however, may be characterised by the out-of-plane dielectric constant $\epsilon_\perp$, which is different from unity $\epsilon_\perp=\epsilon_\perp^{bil}>1$. 
In the following, we demonstrate that the first case can not explain the experimental observables, and hence, it confirms that the BL is polarised in an out-of-plane direction. 

Let us consider the case $\epsilon_\perp=1$.
Then, the excitonic spectrum of the $s$ states can be calculated using the dimensionless eigenvalue equation (with the same notations for $\epsilon$ and $b$ from the previous section)
\begin{equation}
\Big\{b^2\frac{1}{\xi}\frac{d}{d\xi}\Big(\xi\frac{d}{d\xi}\Big)+
bv_{bil}(\xi)+\epsilon\Big\}\psi(\xi)=0,
\end{equation} 
with the dimensionless electrostatic potential between an electron and a hole in the BL
\begin{widetext}
\begin{align}
\label{eq:potential}
v_{bil}(\xi)=2\varepsilon\int_0^\infty dx J_0(x\xi)
\frac{0.932(1-\delta)\left(e^{2xl}[(1-\delta)\varepsilon x+1]-[\varepsilon x(1-\delta)+\delta]\right)
+0.0681-\delta)^2e^{xl}}
{e^{2xl}[(1-\delta)\varepsilon x+1]^2-[\delta+\varepsilon x(1-\delta)]^2}.
\end{align}
\end{widetext}
Here, $J_0(x)$ is the Bessel function of the first kind, $\delta=(\varepsilon-1)/(\varepsilon+1)\approx 0.64$, and and $l=L\varepsilon/r_0\approx 0.56$. 
In the latter, we used $L=6.44\mbox{\AA}$ for MoSe$_2$ BL from HQ Graphene. 
Note that the $v_{bil}(\xi)$ potential is composed of two components: intra- (the term with 0.932 multiplier) and interlayer (the term with 0.068 multiplier). 
One can see that the intralayer term is dominant at a large $l$ distance between the layers, whereas the contribution of the interlayer term decays exponentially with this distance $\propto \exp(-xl)$.
Using numerical solution of the eigenvalue problem with potential $v_{bil}(\xi)$, we obtain
$E_1=-235 $\,meV, $E_2=-68$\, meV, $E_3=-31$\, meV, $E_4=-18$\, meV, and $E_5=-11$\, meV.
The calculated energy positions of the excitonic states in the BL are presented in Fig.~\ref{fig:spectrum}.
Note that the binding energies of the consecutive $s$ excitons in the BL are slightly larger compared to their ML counterparts.
This can be explained by the fact that the effective dielectric constant for the BL is smaller than that for the ML case. 
However, the calculated difference between the energies of the 1$s$ and 2$s$ excitons, $|\Delta E_{12}|=167$\,meV, is significantly larger as compared to the experimental value, $|\Delta E^{exp}_{12}|=118\pm6$\,meV. 
Therefore, the model with an out-of-plane dielectric constant, $\epsilon_\perp=1$ fails in the BL.

Consequently, we calculate the excitonic spectrum in the BL assuming that $\epsilon_\perp=7.7$ \cite{Laturia2018}, which is the value that describes the static dielectric constant in the MoSe$_2$ BL ~\cite{Laturia2018}. 
The resulting eigenvalue equation in dimensionless coordinate $\xi=\rho\varepsilon\sqrt{\epsilon_\perp}/r_0$ becomes 
\begin{equation}
\Big\{b^2\epsilon_\perp \frac{1}{\xi}\frac{d}{d\xi}\Big(\xi\frac{d}{d\xi}\Big)+
b\widetilde{v}_{bil}(\xi)+\epsilon\Big\}\psi(\xi)=0,
\end{equation} 
with the new electrostatic potential, $\widetilde{v}_{bil}(\xi)$, which is obtained from Eq.~(\ref{eq:potential}) 
by replacing $\delta\rightarrow \widetilde{\delta}=(\varepsilon-\sqrt{\epsilon_\perp})/(\varepsilon+\sqrt{\epsilon_\perp})\approx 0.24$. 
The dimensionless energy parameter remains the same as in the previous case.
The eigenvalues of this new equation provide the following spectrum of excitons
$E_1=-177 $\,meV, $E_2=-53$\, meV, $E_3=-26$\, meV, $E_4=-17$\, meV, and $E_5=-12$\, meV.
The calculated energy positions of the excitonic states in the BL with $\epsilon_\perp=7.7$ are presented in Fig.~\ref{fig:spectrum}.
Note that the new energy separation $|\Delta E_{12}|=124$\,meV agrees nicely with the experimental value ($|\Delta E^{exp}_{12}|=118\pm6$\,meV).

To conclude, we demonstrate that the excitation spectrum of excitons in the MoSe$_2$ ML can be properly reproduced using the RK potential with the approach of infinitely thin ML. 
However, the Rytova-Keldysh potential can not be applied to describe the spectrum of the intralayer excitons in the BL as a result of the more complex structure of the BL crystal. 
We derive the electrostatic potential in the BL by taking into account the geometry and the dielectric constant $\epsilon_\perp$ of the BL perpendicular to the layers’ planes. 
The spectrum of excitons, based on the new potential in the MoSe$_2$ BL is in very good agreement with the experimental data.
Our results indicate that the study of excitons in thin layers of \mbox{S-TMDs}, beyond the ML limit, is much more complicated and requires taking into account a realistic thickness of the BL and its dielectric response.

\section{$g$-factors of exciton $ns$ states}
\label{sec:gfactors}
The $g$-factors of excitonic $ns$ states, $g^X_{ns}$, can be calculated using the dispersions of the band-to-band transitions, $g^X(\mathbf{k})$, in the vicinity of the K$^+$ point, along with the exciton wave functions, $|\psi^X_{ns}(\mathbf{k})|$ (obtained from our aforementioned $\mathbf{k\cdot p}$ calculations), following Ref.~\cite{Chen2019}
\begin{equation}
\label{eq:g_ns}
g^X_{ns} = \int_{\mathbf{k}} |\psi^X_{ns}(\mathbf{k})|^2 g^X(\mathbf{k}) d\mathbf{k}
\end{equation}
with
\begin{equation}
g^X(\mathbf{k}) = \pm2 ( g_{c(+1)}(\mathbf{k}) - g_{v(-1)}(\mathbf{k}) ), 
\end{equation}
where the sign is defined by the optical selection rules at K$^{\pm}$ points. $g_{c(+1)}(\mathbf{k})$ and $g_{v(-1)}(\mathbf{k})$ are the $g$-factors of the spin-split subbands (Bloch states), $\ket{m\mathbf{k}}$, involved in the excitonic transition, evaluated as
\begin{equation}
g_m(\mathbf{k}) = L_m^z(\mathbf{k}) + S_m^z(\mathbf{k}). 
\end{equation}
The $z$ component of the orbital angular momentum of a Bloch state is calculated from the bands--summation formula~\cite{Wozniak2020}
\begin{equation}
L_m^z(\mathbf{k}) =  \frac{1}{i m_0} \sum_{l=1, l \neq m}^N 
\frac{p^x_{ml}(\mathbf{k}) p^y_{lm}(\mathbf{k}) - p^y_{ml}(\mathbf{k}) p^x_{lm}(\mathbf{k})}{\varepsilon_m(\mathbf{k}) - \varepsilon_l (\mathbf{k})}, 
\end{equation}
where $m_0$ is the free electron mass, $p^{x,y}_{ml}(\mathbf{k})$ are components of the momentum operator matrix elements, $\varepsilon_m (\mathbf{k})$ are the band energies, and the summation runs over all $N$ states in the basis set. 
The elements $p^{x,y}_{ml}(\mathbf{k})$ were obtained from density perturbation theory calculations \cite{Gajdos2006}. 
In order to converge $L_m^z(\mathbf{k})$ up to 0.1, 480 bands per formula unit were taken into account. 
The spin angular momentum $S_m^z(\mathbf{k})=\pm1$ for the considered bands.
We would like to emphasize that the calculated values of the band $g$-factors for spin-orbit split subbands at the K$^\pm$ points are discussed in the SM.

\begin{figure}[!t]
		\subfloat{}%
		\centering
		\includegraphics[width=1 \linewidth]{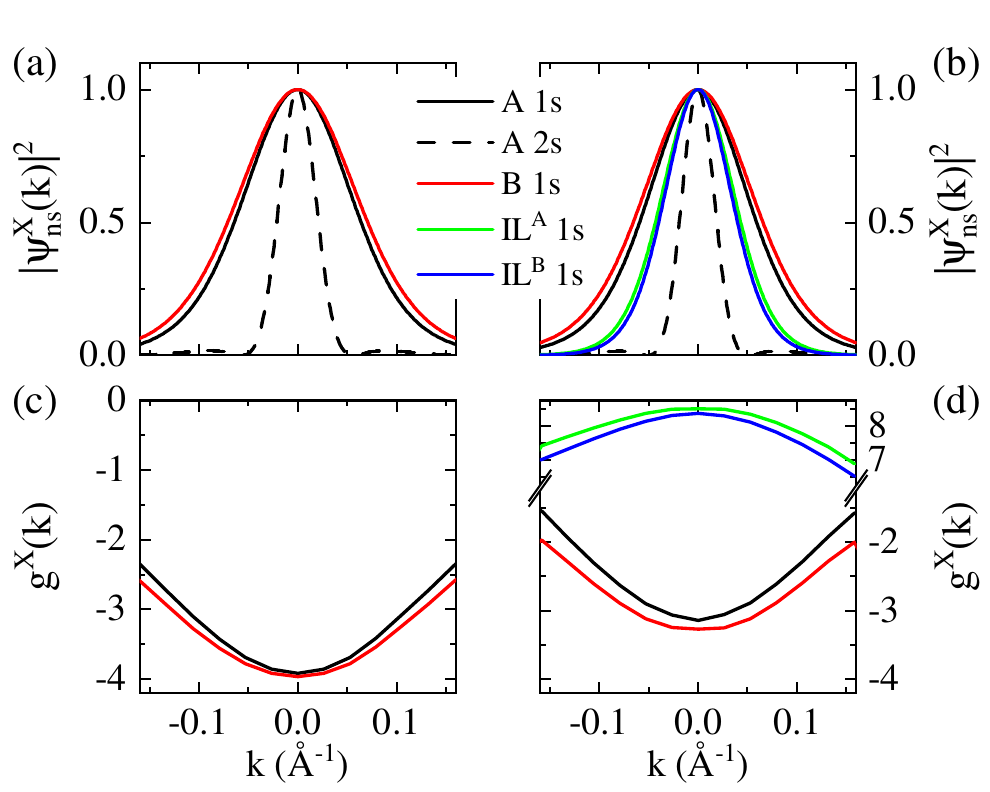}
    	\caption{Normalized squared moduli of exciton wave functions calculated from $\mathbf{k\cdot p}$ model for (a) ML and (b) BL MoSe$_2$. Dispersions of exciton $g$-factors around K$^+$ point from DFT calculations for (c) ML and (d) BL.}
		\label{fig:theory}
\end{figure}

The calculated exciton wave functions squared (see the SM for details) and $g$-factor dispersions in $k$-space around K$^+$ point, that enter Eq.~(\ref{eq:g_ns}), are presented in Fig.~\ref{fig:theory}. 
The widths of the wave functions of the A and B excitons are similar in ML and BL, with a greater spread of $|\psi^B_{1s}(\mathbf{k})|^2$ than $|\psi^A_{1s}(\mathbf{k})|^2$, due to a larger effective mass of the B exciton. 
The opposite is observed for the interlayer excitons. 
The $g^X(\mathbf{k})$ dependence has a positive curvature in both the ML and BL structures, with a smaller $|g^X(\textrm{K})|$ in BL. 
This is caused by the reduction of the bands $g$-factors in BL versus ML (see the SM). 
As a result, the magnitudes of $g^X_{ns}$ in BL are reduced with respect to ML. 
Furthermore, a stronger localisation of the 2$s^\textrm{A}$ state around the K point leads to the increase of their $g$-factor magnitudes, in agreement with previous theoretical and experimental findings~\cite{Chen2019,Raiber2022}. 
Due to the different signs of valence and conduction Bloch states $g$-factors involved in IL$^\textrm{A}$ and IL$^\textrm{B}$ excitons, their $g$-factors are positive and exhibit a negative curvature.

The resulting trends of the calculated $g$-factors are in good agreement with the experimental values, as presented in Tab.~\ref{tab:g-factors}. 
Particularly, the experimentally observed large increase of the $g$-factors of 2$s{^\textrm{A}}$ excitons in ML and BL can not be explained either by the underestimation of the band gap by DFT~\cite{Wozniak2020}, or by the in-plane mechanical strain that might be present in the samples \cite{FariaJunior2022}. 
Further theoretical investigations are required, which are beyond the scope of this study.

\section{Summary \label{sec:summary}}
The excitonic properties in high-quality ML and BL of molybdenum diselenide (MoSe$_2$) encapsulated in hBN flakes were investigated both theoretically and experimentally. 
We determined the $g$-factors of the intralayer A and B excitons and of the interlayer excitons in MoSe$_2$ ML and BL using the RC experiment performed in out-of-plane magnetic fields up to 10~T and first-principle calculations.
The experimental ladder of excitonic $s$ states in the ML was reproduced using the $\mathbf{k\cdot p}$ model with the Rytova-Keldysh potential.
Furthermore, we demonstrated that analogous calculations for the BL require taking into account the out-of-plane dielectric response of the MoSe$_2$ BL, which is neglected for the ML.
Finally, we have explained the values and signs of all observed $g$-factors using a combined $\mathbf{k\cdot p}$ and DFT approach.
Our results manifest that the excitonic physics in S-TMD BLs is more complex than that in the ML case because of the presence of VB hybridization and the non-infinite size of the BL.

\section{Methods \label{sec:method}}
\subsection{Sample and experimental setup}
The investigated MoSe$_2$ thin layers and hBN ﬂakes were fabricated by two-stage PDMS-based mechanical exfoliation of the bulk crystal. 
Initially, the hBN thin flakes were exfoliated onto a 90 nm SiO$_2$/Si substrate and annealed at 200$^\circ$C.
That non-deterministic approach provides the best quality of the substrate surface. Subsequent layers were transferred deterministically using a microscopic system equipped with a $x$-$y$-$z$ motorised positioners. 
The assembled structures were annealed at 160$^\circ$C for 1.5 hour in order to ensure the best layer-to-layer and layer-to-substrate adhesion and to eliminate a substantial portion of air pockets apparent at the interfaces between the constituent layers.

Low-temperature micro-magneto-experiments of reflectance contrast (RC) were performed in Faraday geometry, $i.e.$ magnetic field oriented perpendicularly to the layers plane. 
Measurements (spatial resolution $\sim$3~$\mu$m) were carried out with the aid of a superconducting coil in magnetic fields up to 10~T using an optical fiber arrangement. 
The sample was placed on top of a $x$-$y$-$z$ piezo-stage kept at $T$=10~K and was illuminated using a 100 W tungsten halogen lamp.
The reflectance signal was dispersed with a 0.75 m focal length monochromator and detected with a liquid-nitrogen-cooled Si-CCD. 
The combination of a quarter-wave plate, a linear polariser, and a Wollastom prism was used to analyse the circular polarisation of signals (the $\sigma^\pm$-polarized light was measured simultaneously).
We define the RC spectrum as $\textrm{RC}(E)=[{\textrm{R}(E)-\textrm{R}_0(E)}]/[{\textrm{R}(E)+\textrm{R}_0}(E)] \times 100\%$, where $\textrm{R}(E)$ and $\textrm{R}_0(E)$ are the reflectance of the sample and of the same structure without the ML or BL of MoSe$_2$, respectively.

\subsection{DFT calculations}
First-principles calculations within the DFT were carried out in the Vienna ab-initio simulation package~\cite{Kresse1996}.
The ionic potentials were described using the projector augmented wave technique~\cite{Kresse1999}. We employed the generalised gradient approximation of the exchange correlation-functional within Perdew-Burke-Ernzerhof parametrization~\cite{Perdew1996}. A cutoff energy for the plane-wave basis and a Monkhorst-Pack $k$-grid for the BZ sampling were set to 500~eV and 12$\times$12$\times$1, respectively. The geometrical structures of ML and BL were defined using the parameters from Ref.~\cite{Berkelbach2013} with a vacuum region of 20~\AA\ in order to avoid spurious interactions between the periodically repeated layers. Spin-orbit coupling (SOC) was taken into account during the calculations.

\section*{Data availability statement}
The data that support the findings of this study are available upon reasonable request from the authors.

\section*{Author Contributions}
\L{}.K., M.B., N.Z., K.O.-P., A.B., and M.R.M. performed the experiments.
A.O.S. performed theoretical calculations based on the $\mathbf{k\cdot p}$ approximation.
T.W. carried out DFT calculations.
M.G. fabricated the sample
K.W. and T.T. grew the hBN crystals. 
M.R.M. supervised the project. 
\L{}.K., A.O.S., T.W., and M.R.M. wrote the manuscript with inputs from the all co-authors.

\section*{Conflicts of interest}
There are no conflicts to declare.

\section{Acknowledgements \label{Acknowledgements}}
We thank Paulo E. Faria Junior for fruitful discussions.
The work has been supported by the National Science Centre, Poland (grant no. 2017/27/B/ST3/00205 and 2018/31/B/ST3/02111). 
A.O.S. acknowledges the support by Czech Science Foundation  (project GA23-06369S).
T.W. acknowledges support from the National Science Centre, Poland (grant no. 2021/41/N/ST3/04516).
K.W. and T.T. acknowledge support from JSPS KAKENHI (Grant Numbers 19H05790, 20H00354 and 21H05233).
DFT calculations were performed with the support of Center for Information Services and High Performance Computing (ZIH) at TU Dresden and in part by PLGrid Infrastructure.

\bibliographystyle{apsrev4-2}
\bibliography{biblio}

\newpage
\onecolumngrid
\setcounter{figure}{0}
\setcounter{section}{0}
\renewcommand{\thefigure}{S\arabic{figure}}
\renewcommand{\thesection}{S\Roman{section}}
\include{SM_arxiv}

\end{document}

%% file: SM_arxiv.tex
	\begin{center}
	{\large{{\bf  \textsc{Supplemental Material}} \\ Analogy and dissimilarity of excitons in monolayer and bilayer of MoSe$_2$}}
\end{center}

\section{Optical selection rules in monolayer and bilayer M\lowercase{o}S\lowercase{e}$_\textrm{2}$}
\label{sec:S1}

\begin{figure}[h]
	\centering
	\includegraphics[width=\linewidth]{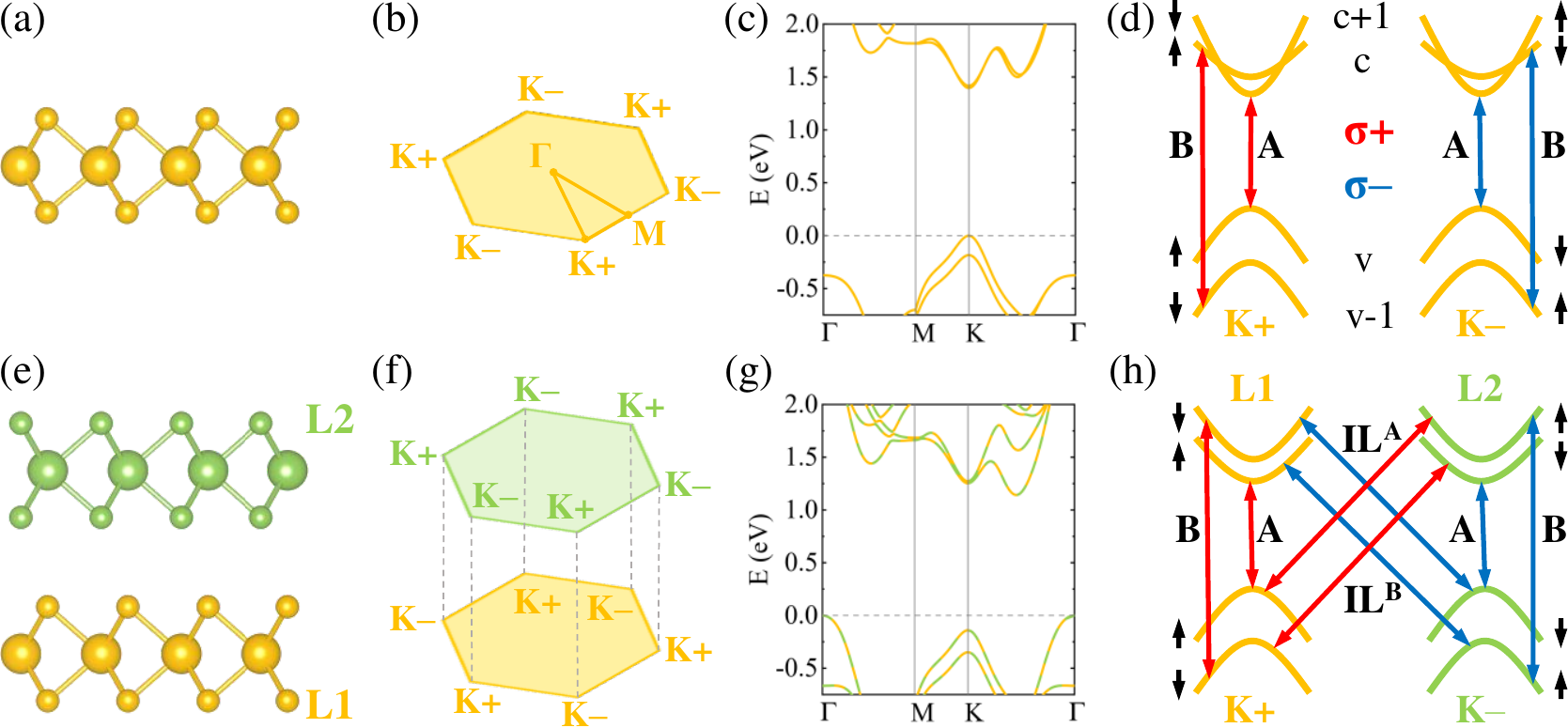}%
	\caption{Side wiev of the atomic structures of (a) ML and (e) BL MoSe$_\textrm{2}$. 2D hexagonal BZs of (b) ML and (f) BL. K$^{\pm}$ points and the $k$-path for band structures are highlighted. Electronic band structures of (c) ML and (g) BL. Bands localized at first (L1) and second (L2) layer are colored orange and green, respectively. Sketch of electronic bands near K$^{\pm}$ points for (d) ML and (h) BL. The spins of bands are marked with black arrows, while the excitonic transitions are marked with red and blue arrows for $\sigma^+$ and $\sigma^-$ polarization of light, respectively. Note that bands at K$^+$ and K$^-$ valleys of L1 and L2 are plotted separately for clarity, while they remain degenerate, as the crystal momenta at K$^+$ of L1 and K$^-$ of L2 are identical. } 
	\label{fig:S1}
\end{figure}

The atomic structure of monolayer (ML) MoSe$_\textrm{2}$, depicted in Fig.~\ref{fig:S1}(a), is non-centrosymmetric. This, with large spin-orbit coupling (SOC) due to the presence of heavy Mo atoms, leads to inequivalency of K$^{\pm}$ points at the corners of hexagonal Brillouin zone (BZ), see Fig.~\ref{fig:S1}(b). Furthermore, the spin degeneracy of the bands is lifted, resulting in spin-split electronic bands (Fig.~\ref{fig:S1}(c)) with opposite spin ordering in the K$^+$ and K$^-$ valleys, as highlighted in Fig.~\ref{fig:S1}(d). As a consequence, the spin-conserving direct transitions between the valence and conduction bands lead to formation of A and B excitons that couple to $\sigma^{\pm}$ light at K$^{\pm}$ valleys, as depicted in Fig.~\ref{fig:S1}(d). 

A bilayer (BL), exfoliated from bulk MoSe$_\textrm{2}$ crystal, consists of two MLs, referred to as L1 and L2, rotated by 60$^\circ$ with respect to each other, as presented in Fig.~\ref{fig:S1}(e). The rotation applies to the BZs of L1 and L2, where the K$^+$ points of L1 are aligned with K$^-$ points of L2 (see Fig.~\ref{fig:S1}(f)). Because the BL atomic structure contains the inversion centre, SOC does not lift the spin degeneracy of the electronic bands. The splittings of the bands visible in the electronic band structures of BL in Fig.~\ref{fig:S1}(g) are caused by the interlayer interactions of L1 and L2. In each pair of doubly degenerate bands, one band is located mostly at L1 (L2) and has spin up (down). Therefore, the spin ordering and optical selection rules for intralayer A and B transitions at $K^{\pm}$ valleys of L1 and L2, shown in Fig.~\ref{fig:S1}(h), are identical to the case of ML. Additionally, two pairs of spin-conserving interlayer excitons with non-zero dipole strength emerge. They involve a hole mainly from L1 (L2) and an electron from L2 (L1) in the opposite K valley and can be viewed as direct in momentum space and indirect in real space. These excitons are referred to as IL$^\textrm{A}$ and IL$^\textrm{B}$, respectively, in analogy to ML case: IL$^\textrm{A}$ (IL$^\textrm{B}$) involve $v$ ($v-1$) bands from K$^+$ (K$^-$) valley of L1 (L2), which implies coupling to $\sigma^+$ ($\sigma^-$) light. Due to the different energetic splitting of the valence and conduction bands, the energies of IL$^\textrm{A}$ and IL$^\textrm{B}$ differ by 170 meV, as estimated from the DFT band energies and $\mathbf{k\cdot p}$ exciton binding energies.

\section{Coulomb potential in multilayer systems}

\subsection{Algorithm of calculation of the Coulomb potential in multilayer systems}

Let us consider the multilayer of S-TMD encapsulated in between hexagonal BN (hBN) flakes. The surface of the bottom hBN flake has a coordinate $z=0$,
and the surface of the top hBN flake has a coordinate $z=L$. The multilayer crystal is presented as a set of $N$ layers of S-TMD, arranged in parallel to the $xy$ plane.
The position of the $j$-th layer is defined by the coordinate $z_j$. We arrange the layers in the following way $0<z_1<z_2<\dots z_N<L$.
We suppose that the multilayer can be polarised mainly in an in-plane direction and weakly in an out-of-plane direction. 
The out-of-plane polarisation properties definitely exist for 3D crystals,
or for crystals with a macroscopical number of layers, however, the polarisation properties of the few-layer crystals remain an open question. 
We suppose that in a few layer systems ($N=2,3,\dots $) the out-of-plane dielectric response can be characterised by phenomenological out-of-plane dielectric constant $\epsilon_\perp$, which can be a function of $N$.

In order to find the potential energy between two charges in the S-TMD few-layers, we solve the following electrostatic problem. We consider the point-like charge $Q$ at the point $\mathbf{r}=(0,0,z')$, where $0<z'<L$ and calculate electric potential in such a system following Ref.~\cite{Cudazzo2011}.
Namely, we consider 3 regions: bottom and top hBN semi-infinite media with potentials $\Phi_1(\boldsymbol{\rho},z)$ and $\Phi_3(\boldsymbol{\rho},z)$, and the space between them with the potential $\Phi_2(\boldsymbol{\rho},z)$.
Here, we introduced the in-plane vector $\boldsymbol{\rho}=(x,y)$.
The Maxwell equations define the form of these potentials.
Using the cylindrical symmetry of the problem, we present the potentials in the form
\begin{equation}
\Phi_j(\boldsymbol{\rho},z)=\frac{1}{(2\pi)^2}\int d^2\mathbf{k} e^{i\mathbf{k}\boldsymbol{\rho}}\Phi_j(\mathbf{k},z).
\end{equation}

Maxwell's equation $\mathrm{div}\mathbf{D}_{1,3}=0$ for the 1-st and 3-rd regions reads
\begin{equation}
-\varepsilon_\parallel\mathbf{k}^2\Phi_{1,3}(\mathbf{k},z)+\varepsilon_\perp\frac{d^2\Phi_{1,3}(\mathbf{k},z)}{dz^2}=0,
\end{equation}
where $\varepsilon_\parallel$ and $\varepsilon_\perp$ are the parallel and perpendicular components of the dielectric constant of hBN, respectively.
The solutions of these equations are
\begin{align}
\Phi_1(\mathbf{k},z)&=Ae^{\kappa z}, \,\,\,\,\quad \text{for}\quad z\leq 0, \\
\Phi_3(\mathbf{k},z)&=Be^{-\kappa z}, \quad \text{for}\quad  z\geq L,
\end{align}
where $\kappa=|\mathbf{k}|\sqrt{\varepsilon_\parallel/\varepsilon_\perp}=
k\sqrt{\varepsilon_\parallel/\varepsilon_\perp}$.

The equation for the potential $\Phi_2(\mathbf{r},z)$ takes a form
\begin{align}
\Delta_\parallel\Phi(\boldsymbol{\rho},z)+\epsilon_\perp\frac{d^2\Phi(\boldsymbol{\rho},z)}{dz^2}=-4\pi Q\delta(\boldsymbol{\rho})\delta(z-z')+4\pi \varrho_\text{ind}(\boldsymbol{\rho},z),
\end{align}
where $\Delta_\parallel=\partial_x^2+\partial_y^2$ is the 2D Laplace operator and we introduced phenomenological out-of-plane dielectric constant $\epsilon_\perp$.
The first term in the r.h.s. of the equation is the charge density of the charge $Q$.
The second term represents the induced charge density $\varrho_\text{ind}(\boldsymbol{\rho},z)$ due to 
the polarisation of layers by charge $Q$. The corresponding charge density $\varrho_\text{ind}(\boldsymbol{\rho},z)$ and the polarisation of the layers $\mathbf{P}(\boldsymbol{\rho},z)$ are coupled as 
$\varrho_\text{ind}(\mathbf{r},z)=\mathrm{div} \mathbf{P}(\boldsymbol{\rho},z)$. 
Following Ref.~\cite{Cudazzo2011}, we present the polarisation in the form
\begin{equation}
\mathbf{P}(\boldsymbol{\rho},z)=\sum_{j=1}^N \delta(z-z_j)\mathbf{P}_\parallel(\boldsymbol{\rho},z_j).
\end{equation}

Using the linear response $\mathbf{P}_\parallel(\boldsymbol{\rho},z_j)=
\chi_{2D}\mathbf{E}_\parallel(\boldsymbol{\rho},z_j)$
we get the expression for the induced charge
\begin{equation}
\varrho_{ind}(\mathbf{r},z)=-\chi_{2D}\sum_{j=1}^N \delta(z-z_j)\Delta_\parallel\Phi_2(\boldsymbol{\rho},z_j).
\end{equation}

Taking Fourier transformation one gets
\begin{align}
\Big[\mathbf{k}^2-\epsilon_\perp\frac{d^2}{dz^2}\Big]\Phi_2(\mathbf{k},z)=4\pi Q\delta(z-z')- 
4\pi\chi_{2D}\mathbf{k}^2\sum_{j=1}^N \delta(z-z_j)\Phi_2(\mathbf{k},z_j).
\end{align}

Integrating this equation in the regions $z\in[z_j-\epsilon, z_j+\epsilon]$ and $z\in[z'-\epsilon,z'+\epsilon]$ 
with infinitesimally small $\epsilon>0$ one gets the following conditions
\begin{align}
\epsilon_\perp\frac{d\Phi_2(\mathbf{k},z)}{dz}\Big|^{z_j+\epsilon}_{z_j-\epsilon}&=2r_0k^2\Phi_2(\mathbf{k},z_j),\\
\epsilon_\perp\frac{d\Phi_2(\mathbf{k},z)}{dz}\Big|^{z'+\epsilon}_{z'-\epsilon}&=-4\pi Q,
\end{align}
where we introduced $r_0=2\pi\chi_{2D}$. $\chi_{2D}$ is the material’s 2D polarisability. Outside of the points $z_j$ and $z'$ we have the equation
\begin{equation}
\Big[\mathbf{k}^2-\epsilon_\perp\frac{d^2}{dz^2}\Big]\Phi_2(\mathbf{k},z)=0
\end{equation}

The general solution for $\Phi_2(\mathbf{k},z)$ can be written in the form
\begin{equation}
\label{eq:potential2}
\Phi_2(\mathbf{k},z)=\Psi_0e^{-K|z-z'|}+\sum_{j=1}^N \Psi_je^{-K|z-z_j|} + \alpha e^{Kz} +\beta e^{-Kz},
\end{equation}
where $\Psi_0$,$\Psi_j$, $\alpha$ and $\beta$ are unknown functions of the parameter $K=k/\sqrt{\epsilon_\perp}$.

The aforementioned boundary conditions together with the equation give the following 
\begin{align}
\label{eq:boundary_conditions_1}
\Psi_j=&-r_0k\Phi_2(\mathbf{k},z_j)/\sqrt{\epsilon_\perp}=-r_0K\Phi_2(\mathbf{k},z_j), \\
\label{eq:boundary_conditions_2}
\Psi_0=&2\pi Q/k\sqrt{\epsilon_\perp}=2\pi Q/\epsilon_\perp K.
\end{align}
Taking the boundary conditions at $z=0$ and $z=L$
\begin{align}
\Phi_1(\mathbf{k},0)=&\Phi_2(\mathbf{k},0), \quad  \frac{\varepsilon_\perp}{\epsilon_\perp}
\frac{d\Phi_1(\mathbf{k},0)}{dz}=\frac{d\Phi_2(\mathbf{k},0)}{dz},  \\
\Phi_3(\mathbf{k},L)=&\Phi_2(\mathbf{k},L), \quad \frac{\varepsilon_\perp}{\epsilon_\perp}
\frac{d\Phi_3(\mathbf{k},L)}{dz}=\frac{d\Phi_2(\mathbf{k},L)}{dz},
\end{align}
we get the following equations
\begin{align}
A=&\Psi_0e^{-Kz'}+\sum_{j=1}^N\Psi_je^{-Kz_j}+\alpha+\beta, \\ 
\varepsilon A=&\Psi_0e^{-Kz'}+\sum_{j=1}^N\Psi_je^{-Kz_j}+\alpha-\beta, \\
Be^{-\kappa L}=&\Psi_0e^{-K(L-z')}+\sum_{j=1}^N\Psi_je^{-K(L-z_j)}+\alpha e^{KL}+\beta e^{-KL},  \\
\varepsilon Be^{-\kappa L}=&\Psi_0e^{-K(L-z')}+\sum_{j=1}^N\Psi_je^{-K(L-z_j)}-\alpha e^{KL}+\beta e^{-KL}, 
\end{align}
where we introduced $\varepsilon=\sqrt{\varepsilon_\perp\varepsilon_\parallel/\epsilon_\perp}$. 
By removing the parameters $A$ and $B$ from these equations, one gets for $\varepsilon\neq1$
\begin{align}
\Psi_0e^{-Kz'}+\sum_{j=1}^N\Psi_je^{-Kz_j}+\alpha+\frac{\beta}{\delta}=0,
\end{align}
\begin{align}
\Psi_0e^{-K(L-z')}+\sum_{j=1}^N\Psi_je^{-K(L-z_j)}+\frac{\alpha e^{KL}}{\delta}+\beta e^{-KL}=0, 
\end{align}
where $\delta=(\varepsilon-1)/(\varepsilon+1)\in(0,1)$. 
Note that for the case $\varepsilon=1$ we get $\alpha=\beta=0$, and therefore
the solution in the second region
\begin{equation}
\Phi_2(\mathbf{k},z)=\Psi_0e^{-K|z-z'|}+\sum_{j=1}^N \Psi_je^{-K|z-z_j|},
\end{equation}
One can verify that the solution for the potential, in this case, can be obtained as a limit $\delta\rightarrow 0$ of the general solution. Therefore, we consider the general case with $\delta\neq0$. 

We solve the general equation in a few steps. First, we write the system of equations in the following form
\begin{align}
\label{eq:alphabeta}
\delta\alpha+\beta=\delta\mathcal{A}, \quad e^{KL}\alpha+\delta e^{-KL}\beta=\delta\mathcal{B},
\end{align}
where we introduced $\mathcal{A}=-\Psi_0e^{-Kz'}-\sum_{j=1}^N\Psi_je^{-Kz_j}$, 
$\mathcal{B}=-\Psi_0e^{-K(L-z')}-\sum_{j=1}^N\Psi_je^{-K(L-z_j)}$.
Solving equations (\ref{eq:alphabeta}) one gets
\begin{align}
\alpha=\frac{\delta[\mathcal{B}e^{KL}-\mathcal{A}\delta]}{e^{2KL}-\delta^2},
 \quad \beta=\frac{\delta e^{KL}[\mathcal{A}e^{KL}-\mathcal{B}\delta]}{e^{2KL}-\delta^2}.
\end{align}
Substituting these solutions into (\ref{eq:potential2}) we express $\Phi_2(\mathbf{k},z)$ 
via $N+1$ parameters $\Psi_0$ and $\Psi_j$. Then, using the latter expression,
we calculate the values for the potential at $z'$ and $z_l,\, l=1,2\dots N$ points
\begin{align}
\label{eq:self_consisted}
\Phi_2(\mathbf{k},z_l)=&\Psi_0e^{-K|z_l-z'|}+\sum_{j=1}^N \Psi_je^{-K|z_l-z_j|} + \alpha e^{Kz_l} +\beta e^{-Kz_l},
\end{align}
which together with (\ref{eq:boundary_conditions_1}) and (\ref{eq:boundary_conditions_2}) give the complete set of equations for the parameters 
$\Psi_0$ and $\Psi_j$.

Note that the system of equations (\ref{eq:boundary_conditions_1}), (\ref{eq:boundary_conditions_2}), (\ref{eq:alphabeta}) and 
(\ref{eq:self_consisted}) are simplified at 
$\epsilon_\perp=1$. We first obtain the solution for this particular case. 
The general solution for arbitrary $\epsilon_\perp$ can be obtained from the previous solution 
by replacing $k\rightarrow K=k/\sqrt{\epsilon_\perp}$, $Q\rightarrow Q/\epsilon_\perp$, and 
$\varepsilon\rightarrow \varepsilon/\sqrt{\epsilon_\perp} $ as one can see from the aforementioned system of equations and the definition of $\varepsilon$ and $K$. 

In the following, we derive the potentials for the case of monolayer (ML) and bilayer (BL) of S-TMDs.

\subsection{Coulomb potential in the monolayer}

We solve the system of equations (\ref{eq:self_consisted}) for the S-TMD monolayer encapsulated between two dielectric flakes with a distance $L$ between them. We consider the case where the charge $Q$ is placed inside the monolayer $z'=z_1$ with a symmetric disposition of the monolayer between dielectric media $z_1=L/2$.
The solution is
\begin{equation}
\Phi_2(\mathbf{k},L/2)=\frac{2\pi Q}{k}\frac{1-\delta e^{-kL}}{1+kr_0-\delta(kr_0-1)e^{-kL}}.
\end{equation}
In the $L\rightarrow\infty$ limit the answer does not depend on $\delta$ and reproduces the case of the potential for 
the monolayer in a vacuum
\begin{equation}
\Phi_2(\mathbf{k},L/2)=\frac{2\pi Q}{k}\frac{1}{1+kr_0}.
\end{equation}
At $L\rightarrow0$ limit it coincides with the expression for the generalised Rytova-Keldysh potential as a function of an in-plane distance $\rho=\sqrt{x^2+y^2}$ \cite{Cudazzo2011, Keldysh1979, Rytova1967}
\begin{align}
\Phi_2(\mathbf{k},0)=\frac{2\pi Q}{k}\frac{1-\delta}{1+\delta+(1-\delta)kr_0}=
\frac{2\pi Q}{k\varepsilon}\frac{1}{1+kr_0/\varepsilon}.
\end{align}
Using the Fourier transform of the latter expression, we restore the Rytova-Keldysh potential~\cite{Rytova1967, Keldysh1979}
\begin{equation}
V_{RK}(\rho)=
\frac{\pi Q}{2r_0}\Big[\text{H}_0\Big(\frac{\rho\varepsilon}{r_0}\Big)-Y_0\Big(\frac{\rho\varepsilon}{r_0}\Big)\Big],
\end{equation}
where $\text{H}_0(x)$ and $Y_0(x)$ are the Struve function and the Bessel function of the second kind.

\subsection{Coulomb potential in the bilayer}

We consider two monolayers positioned at $z_1=\xi$ and $z_2=L-\xi$. We calculate the in-plane potential 
$\Phi_2(\mathbf{k},z)$ for the cases of $z=\xi$ an $z=L-\xi$  with the charge
$Q$ placed in $z'=z_1=\xi$ and then consider the limit $\xi\rightarrow0$ for simplicity. The answer is
\begin{align}
\label{eq:phi_1}
\Phi_2(\mathbf{k},0)=&\frac{2\pi Q}{k}\frac{(1-\delta)\left(e^{2kL}[(1-\delta)kr_0+1]-[kr_0(1-\delta)+\delta]\right)}
{e^{2kL}[(1-\delta)kr_0+1]^2-[\delta+kr_0(1-\delta)]^2},\\
\label{eq:phi_2}
\Phi_2(\mathbf{k},L)=&\frac{2\pi Q}{k}\frac{(1-\delta)^2e^{kL}}{e^{2kL} [(1-\delta)kr_0+1]^2-[\delta +kr_0(1-\delta)]^2}.
\end{align}
The limit case $L\rightarrow0$ gives the result for the monolayer with twice larger screening length $r_0\rightarrow 2r_0$
\begin{align}
\Phi_2(\mathbf{k},0)=\Phi_2(\mathbf{k},L\rightarrow 0)=
\frac{2\pi Q}{k\varepsilon}\frac{1}{1+2kr_0/\varepsilon}, 
\end{align}
as it should be. This result is apparent, since in the case of zero distance between layers, 
the electrostatic response of the bilayer is twice larger because of the geometry of the system. 

For $L\rightarrow\infty$ we have $\Phi_2(\mathbf{k},L\rightarrow\infty)=0$, therefore, the charges in different layers do not interact with each other in this limit. 
The potential between charges at the same layer in this limit takes the following form
\begin{align}
\Phi_2(\mathbf{k},0)=\frac{2\pi Q}{k\varepsilon_{eff}}\frac{1}{1+kr_0/\varepsilon_{eff}},
\end{align}
where $\varepsilon_{eff}=(\varepsilon+1)/2$. 
It reproduces the result for the monolayer deposited on the hBN substrate.

As a summary, we note that the electrostatic potential in the bilayer is a complex function of the parameter $L$. 
At large distances $L\gg r_0$ the role of the second layer becomes negligible and the electrostatic potential 
within the layer can increase due to the different dielectric coefficients of the environments above and below of 
the considered monolayer of the bilayer. Namely, the potential increases when the dielectric constant of 
the top environment $\varepsilon_\text{top}$ is smaller than the dielectric constant of the bottom hBN substrate $\varepsilon_\text{bott}=\varepsilon_\text{hBN}=4.5$. For the opposite limit, $L\ll r_0$ the screening of 
the second layer dominates and can suppress the effect of the small $\varepsilon_\text{top}$ dielectric constant. 
Therefore, one should estimate that the spectrum of the excitons in the bilayer can be sensitive to the values of 
$L$ and $r_0$ parameters.

\section{Eigenvalue and eigenfunctions problem for intralayer excitons in few-layer systems}
    
To investigate the ML and the BL of S-TMDs, we consider the electron and hole excitations in $\mathbf{k\cdot p}$ approximation, derive the corresponding eigenvalues and eigenfunctions problem for the intravalley excitons \cite{Kormanyos2015, Slobodeniuk2019}. 
For the BL, we consider two cases $\epsilon_\perp=1$, 
and $\epsilon_\perp=\epsilon_\perp^{bil}$, where $\epsilon_\perp^{bil}$ is  
the out-of-plane dielectric constant of the bilayer, in order to understand the role of the 
phenomenological parameter $\epsilon_\perp$ for the considered problem.    

\subsection{Monolayer} 

The spectrum of excitons with reduced mass $\mu$ in the S-TMD monolayer can be obtained by solving the equation
\begin{equation}
\Big\{\frac{\hbar^2}{2\mu}\nabla_\parallel^2-V_{RK}(\rho)+E\Big\}\psi(\rho)=0, 
\end{equation} 
where  
\begin{equation}
V_{RK}(\rho)=-\frac{\pi e^2}{2r_0}\Big[\text{H}_0\Big(\frac{\rho\varepsilon}{r_0}\Big)-
Y_0\Big(\frac{\rho\varepsilon}{r_0}\Big)\Big].
\end{equation}
Introducing the dimensionless parameters $\xi=\rho\varepsilon/r_0=\rho/r_0^*$ and $E=(\mu e^4/2\hbar^2\varepsilon^2)\epsilon=Ry^*\epsilon$ one rewrites the equation for the case of $s$-type excitons in the following dimensionless form
\begin{equation}
\Big\{b^2\frac{1}{\xi}\frac{d}{d\xi}\Big(\xi\frac{d}{d\xi}\Big)+b v_{RK}(\xi)+\epsilon\Big\}\psi(\xi)=0,
\end{equation} 
where $b=\hbar^2\varepsilon^2/\mu e^2r_0$ and 
\begin{equation}
v_{RK}(\xi)=\pi[\text{H}_0(\xi)-Y_0(\xi)].
\end{equation}
The scale and mass parameters for the MoSe$_2$ monolayer are quite broad: 
$r_0=51.7\mbox{\AA}$, $\mu=0.27m_0$ \cite{Berkelbach2013}; 
$r_0=39\mbox{\AA}$, $\mu=0.35m_0$ \cite{Goryca2019};
$\mu=0.44m_0$ \cite{Molas2019_1};
$\mu=0.27m_0$ \cite{Kormanyos2015} where $m_0$ is electron's mass.
In our calculation, we use the following numbers $\varepsilon=4.5$, $r_0=51.7\mbox{\AA}$, $\mu=0.44m_0$.
We obtain $b\approx 0.47$.
The numerical solution of the equation for this value of $b$
gives $\epsilon_1=-0.723$, $\epsilon_2=-0.213$, $\epsilon_3=-0.101$, $\epsilon_4=-0.058$, and $\epsilon_5=-0.038$ for the first excitonic states.
Using the value of the Rydberg energy $Ry^*\approx 295.5$\,meV, 
we find the relative energies of the ground (1$s$) and the four excited (2$s$, 3$s$, 4$s$, and 5$s$) states equal to $E_1=-214$\,meV, $E_2=-63$\,meV, $E_3=-30$\,meV, $E_4=-17$\,meV, $E_5=-11$\,meV, respectively. 
Note that the energies of 1s $E_{1,ex}=E_\text{g}+E_1$ and 2s $E_{2,ex}=E_\text{g}+E_2$ exciton resonances are hard to predict, since the bandgap $E_\text{g}$ is renormalized by the Coulomb interaction, and depends on dielectric constant $\varepsilon$. 
The calculation of such a bandgap shift requires additional numerical investigation. For our study, we exclude this shift by considering the    
energy distance between the 1$s$ and 2$s$ excitons $\Delta E_{12}=E_{1, ex}-E_{2, ex}=
(E_\text{g}+E_1)-(E_\text{g}+E_2)$. 
Therefore, we have $|\Delta E_{12}|=151$\,meV, which is very close to the experimentally observed value is $|\Delta E^{exp}_{12}|=152$\,meV.

To have a full picture we calculate the wave functions of the 1s and 2s excitonic states in the monolayer. 
Then, one can fit these wave functions to the hydrogen type 1s and 2s states 
\begin{align}
\psi_{1s}(\boldsymbol{\rho})=&\frac{1}{\sqrt{2\pi}}\beta e^{-\beta \rho/2}, \\
\psi_{2s}(\boldsymbol{\rho})=&\frac{1}{\sqrt{2\pi}}\frac{\beta^2}{\sqrt{6\alpha^2-4\alpha\beta+\beta^2}}
(1-\alpha \rho)e^{-\beta \rho/2}.
\end{align}
Both functions can be written in the momentum space
\begin{align}
\psi_{ns}(\boldsymbol{\rho})=&\frac{1}{2\pi}\int d^2\mathbf{k} \,e^{i\mathbf{k}\boldsymbol{\rho}}\psi_{ns}(\mathbf{k}), \\
\psi_{ns}(\mathbf{k})=&\frac{1}{2\pi}\int d^2\boldsymbol{\rho}\, e^{-i\mathbf{k}\boldsymbol{\rho}}\psi_{ns}(\boldsymbol{\rho}).
\end{align}
Both functions in momentum space are the functions of the absolute values of $|\mathbf{k}|=k$, 
due to $|\boldsymbol{\rho}|=\rho$ dependence of the corresponding functions in the coordinate space
$\psi_{ns}(\mathbf{k})=\psi_{ns}(k)$,  $\psi_{ns}(\boldsymbol{\rho})=\psi_{ns}(\rho)$. Then we have
\begin{equation}
\psi_{ns}(k)=\int_0^\infty d\rho \rho J_0(k\rho) \psi_{ns}(\rho),
\end{equation}
where $J_0(x)$ is the zeroth Bessel function of the first kind.   
The integration gives
\begin{align}
\psi_{1s}(k)=&\frac{2 \sqrt{\frac{2}{\pi }} \beta ^2}{\left(\beta ^2+4 k^2\right)^{3/2}}, \\
\psi_{2s}(k)=&\frac{2 \sqrt{\frac{2}{\pi }} \beta ^2 \left(\beta ^2 (\beta -4 \alpha )+4 k^2 (2 \alpha +\beta )\right)}{\sqrt{6 \alpha ^2-4 \alpha  \beta +\beta ^2} \left(\beta ^2+4 k^2\right)^{5/2}}.
\end{align}
The latter expressions define the size of the wave functions in momentum space. 
Namely, comparing the $k$-dependent and $k$-independent terms in  the denominator, i.e.,  
$\beta^2=4k^2$, one gets the value of $k_0$ that satisfies the latter equation. 
For both 1s and 2s hydrogen-like functions it is $k_0=\beta/2$, which is nothing but 
the parameter which defines the exponential decay of the 
wave-function in coordinate space $\psi_{ns}(\rho)\sim e^{-\beta \rho/2}$. 
However, the numerical analysis of wave functions in the Rytova-Keldysh potential demonstrates that 
the numerically obtained 1s and 2s wave functions can not be fitted well with the Hydrogenic ones.
Therefore, we calculate numerically the Fourier transform of the dimensionless wave-function $\psi_{ns}(\xi=\rho/r_0^*)$
\begin{align}
\psi_{ns}(k)=\int_0^\infty d\rho \rho J_0(k\rho)\psi_{ns}(\rho)=
(r_0^*)^2 \int_0^\infty d\xi \xi J_0(kr_0^*\xi)\psi_{ns}(\xi)
\end{align}
and estimate the size of the excitonic wave function from its shape in momentum space.
We fit the obtained expressions for $k$-dependent 1s and 2s states by two- and three-parametric 
functions similar to Hydrogen ones. The 1s state can be approximated by the function   
\begin{equation}
\psi_{1s}(k)/\psi_{1s}(0)=\frac{1}{\left(0.312656 (kr_0^*)^2+1\right)^{2.1987}},
\end{equation} 
with $\psi_{1s}(0)=0.581476 r_0^*$.
The 2s wave function can be approximated as 
\begin{equation}
\psi_{2s}(k)/\psi_{2s}(0)=\frac{1-2.36996 (kr_0^*)^2}{(1.02795 (kr_0^*)^2+1)^{3.28023}},
\end{equation}   
with $\psi_{2s}(0)=1.68132 r_0^*$.
Using the aforementioned we can estimate the following sizes of these wave-functions
\begin{equation}
k_{1s}\approx \frac{1.78841}{r_0^*}, \quad k_{2s}\approx \frac{0.98631}{r_0^*}, 
\end{equation}
where $r_0^*\approx 11.5\mbox{\AA}$.

We estimate the reduced mass $\mu=0.515 m_0$ of B-excitons using the experimental value $|\Delta E_{12}|=158$\,meV.  
To do this, we calculate numerically the difference $|\Delta E_{12}(\mu)|$ as a function of the parameter $\mu$, and solve the equation $|\Delta E_{12}(\mu)|=158$\,meV. Then, using the obtained value of $\mu$ we estimate and fit the 1s and 2s states in the momentum space.
Then the corresponding wave functions for 1s and 2s states of B-excitons in the momentum space are 
\begin{align}
\psi_{1s}(k)/\psi_{1s}(0)=\frac{1}{(0.253136 (kr_0^*)^2+1)^{2.22737}},
\end{align}
with $\psi_{1s}(0)=0.527605 r_0^*$, and 
\begin{align}
\psi_{2s}(k)/\psi_{2s}(0)=\frac{1- 1.91068(kr_0^*)^2}{\left(0.809057(kr_0^*)^2+1\right)^{3.3087}},
\end{align}
with $\psi_{2s}(0)=1.49775 r_0^*$. Using the aforementioned we can estimate the following sizes of these wave-functions
\begin{equation}
k_{1s}\approx \frac{1.98757}{r_0^*}, \quad k_{2s}\approx \frac{1.11176}{r_0^*}.
\end{equation}

\subsection{Bilayer: $\epsilon_\perp=1$} 

\begin{figure}[b!]
	\centering
	\includegraphics[width=0.3\linewidth]{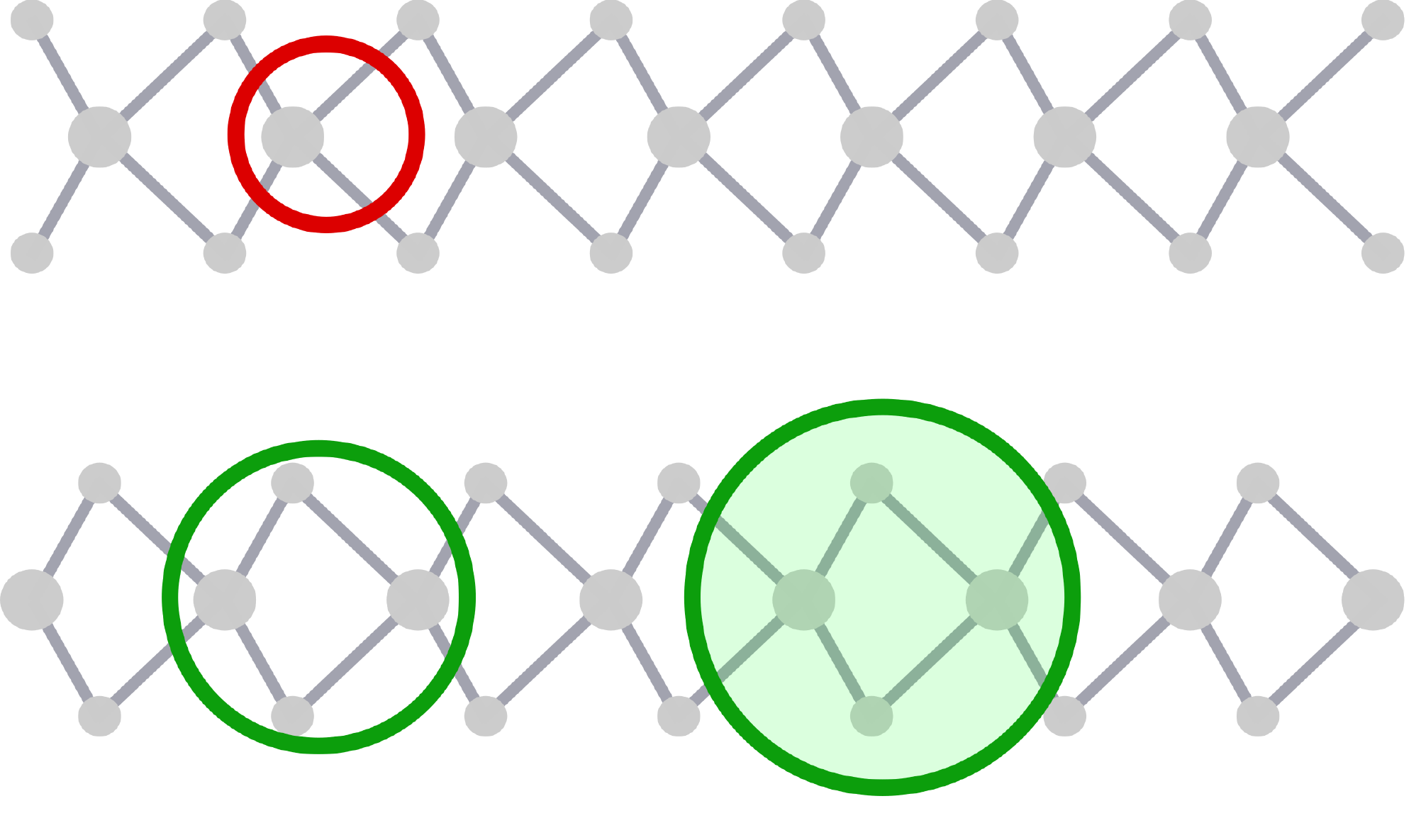}%
	\caption{The schematic picture of the intralayer exciton in the S-TMD BL in the K$^+$ point. 
	The filled circle represents the conduction band electron localized in the bottom layer. 
	Two empty circles represent the hole quasiparticle in the bilayer, localized mainly in the bottom layer. 
	The size of the circles denotes the amount of charge of quasiparticles redistributed in each layer. 
	The colors of the circles encode the polarizations of photons which can be coupled to the exciton: 
	green for $\sigma^+$ and red for $\sigma^-$. The latter means that the intralayer exciton of the bilayer 
	in the K$^+$ point is mainly active in $\sigma^+$ polarization and weakly active in $\sigma^-$ polarization.} 
	\label{fig:bilayer_2022}
\end{figure}
To calculate the binding energies of A-excitons in the bilayer, one needs to take into account the admixture of the
valence band states from the opposite valleys in different layers (see Ref.~\cite{Molas2017Nanoscale, Slobodeniuk2019} for details). Due to this admixture, the charge of the hole quasiparticle is redistributed between two layers, see Fig.~\ref{fig:bilayer_2022}.
It modifies the Coulomb potential between electrons and holes in the bilayer. 
Taking into account the results of Ref.~\cite{Slobodeniuk2019} we obtain the following 
value for the charges of the hole, which belongs to the same plane as an electron 
\begin{equation}
Q_{in}=\frac{|e|}{2}\Big(1+\frac{\Delta_v}{\sqrt{\Delta_v^2+4t^2}}\Big)\approx 0.932|e|,
\end{equation}
and for a charge of the hole, which belongs to the opposite plane as an electron
\begin{equation}
Q_{opp}=\frac{|e|}{2}\Big(1-\frac{\Delta_v}{\sqrt{\Delta_v^2+4t^2}}\Big)\approx 0.068|e|.
\end{equation} 
Here, we used the numbers $\Delta_v=182$\,meV and $t=53$\,meV \cite{Gong2013}.   
Therefore, the Coulomb potential between an electron and hole quasiparticle is a superposition of the intra- and interlayer 
potentials 
\begin{align}
V_{bil}(k)=-\frac{2\pi e^2}{k} \frac{0.932(1-\delta)\left(e^{2kL}[(1-\delta)kr_0+1]-[kr_0(1-\delta)+\delta]\right)
+0.068(1-\delta)^2e^{kL}}
{e^{2kL}[(1-\delta)kr_0+1]^2-[\delta+kr_0(1-\delta)]^2},
\end{align}
where we use $\delta\approx 0.64$, $L=6.44\mbox{\AA}$ (data from HQ Graphene)
and $r_0=51.7\mbox{\AA}$.

The coordinate-dependent potential then reads
\begin{align}
V_{bil}(\rho)=-e^2\int_0^\infty dk J_0(k\rho)
\frac{0.932(1-\delta)\left(e^{2kL}[(1-\delta)kr_0+1]-[kr_0(1-\delta)+\delta]\right)
+0.068(1-\delta)^2e^{kL}}
{e^{2kL}[(1-\delta)kr_0+1]^2-[\delta+kr_0(1-\delta)]^2}
\end{align}
Taking into account that the effective masses of the top and bottom valence bands in MoS$_2$ are  close  
\cite{Kormanyos2015} we assume that the effective mass of the new hole quasiparticles is the same as in ML, and hence the 
reduced mass $\mu=0.44m_0$ coincides with the one in monolayer.    
Then the energies of 1s and 2s A excitons in the bilayer can be found from the equation 
\begin{equation}
\Big\{\frac{\hbar^2}{2\mu}\nabla_\parallel^2-V_{bil}(\rho)+E\Big\}\psi(\rho)=0.
\end{equation}
Introducing again the dimensionless parameters of energy $\epsilon=E/Ry^*$ and distance 
$\xi=\rho\varepsilon/r_0$ we obtain the following equation
\begin{equation}
\Big\{b^2\frac{1}{\xi}\frac{d}{d\xi}\Big(\xi\frac{d}{d\xi}\Big)+
bv_{bil}(\xi)+\epsilon\Big\}\psi(\xi)=0,
\end{equation} 
where 
\begin{align}
\label{eq:v_bil}
v_{bil}(\xi)=2\varepsilon\int_0^\infty dx J_0(x\xi)
\frac{0.932(1-\delta)\left(e^{2xl}[(1-\delta)\varepsilon x+1]-[\varepsilon x(1-\delta)+\delta]\right)
+0.068(1-\delta)^2e^{xl}}
{e^{2xl}[(1-\delta)\varepsilon x+1]^2-[\delta+\varepsilon x(1-\delta)]^2}
\end{align}
and $l=L\varepsilon/r_0\approx 0.56$.  
Note that the potential is a sum of intra- $v_{1,bil}(x)$ and interlayer $v_{2,bil}(x)$ 
contributions
\begin{equation}
v_{bil}(\xi)=\int_0^\infty dx J_0(x\xi) [v_{1,bil}(x)+v_{2,bil}(x)], 
\end{equation}
where 
\begin{align}
v_{1,bil}(x)=&2\varepsilon
\frac{0.932(1-\delta)\left(e^{2xl}[(1-\delta)\varepsilon x+1]-[\varepsilon x(1-\delta)+\delta]\right)}
{e^{2xl}[(1-\delta)\varepsilon x+1]^2-[\delta+\varepsilon x(1-\delta)]^2},\\
v_{2,bil}(x)=&2\varepsilon
\frac{0.068(1-\delta)^2e^{xl}}
{e^{2xl}[(1-\delta)\varepsilon x+1]^2-[\delta+\varepsilon x(1-\delta)]^2}.
\end{align}
They correspond to the coordinate-dependent terms 
\begin{equation}
v_{j,bil}(\xi)=\int_0^\infty dx J_0(x\xi)v_{j,bil}(x), \quad j=1,2,
\end{equation}
as functions of dimensionless coordinate $\xi=r\varepsilon/r_0$. 
One can observe that the intralayer term contains the long $\sim 1/x$ tail which is 
responsible for the singular behaviour of $v_{1,bil}(\xi)$ at the origin $\xi=0$. In order to take into account this singular term exactly, we extract the $x\rightarrow \infty$ limit  
\begin{equation}
v^{lim}_{1,bil}(x)=\frac{0.932\times 4\varepsilon}{2\varepsilon x+\varepsilon+1}
\end{equation}     
from the potential $v_{1,bil}(\xi)$ and rewrite it in the form  
\begin{align}
v_{1,bil}(x)=v^{lim}_{1,bil}(x)+[v_{1,bil}(x)-v^{lim}_{1,bil}(x)]=
v^{lim}_{1,bil}(x)+\delta v_{1,bil}(x).
\end{align}
The correction $\delta v_{1,bil}(x)$ is a bounded function, which exponentially 
decays in $x\rightarrow \infty$ limit. Therefore, it gives a small and non-singular contribution to the coordinate-dependent potential $\delta v_{1,bil}(\xi)$ at the origin $\xi=0$. The interlayer term $v_{2,bil}(x)\propto \exp(-xl)$ is also exponentially suppressed at $x\rightarrow \infty$, and, hence, its coordinate-dependent contribution $v_{2,bil}(\xi)$ is regular at the origin. For the numerical 
calculation of the spectrum and corresponding wave functions of the excitonic states in the bilayer we present 
the full potential in the form 
\begin{align}
v_{bil}(\xi)=v_{1,bil}^{lim}(\xi)+\delta v_{1,bil}(\xi)+v_{2,bil}(\xi),
\end{align}
considering the exact expression for the first term 
\begin{align}
v_{1,bil}^{lim}(\xi)=\int_0^\infty dx J_0(x\xi)v_{1,bil}^{lim}(x)=
0.932\pi 
\Big[\text{H}_0\Big(\frac{\xi(\varepsilon+1)}{2\varepsilon}\Big)-
Y_0\Big(\frac{\xi(\varepsilon +1)}{2\varepsilon}\Big)\Big]=
0.932v_{RK}\Big(\frac{\xi(\varepsilon+1)}{2\varepsilon}\Big).
\end{align}

Solving numerically the eigenvalue problem with $v_{bil}(\xi)$ potential, we obtain $\epsilon_1=-0.794$, $\epsilon_2=-0.229$, $\epsilon_3=-0.105$, $\epsilon_4=-0.0598$, and $\epsilon_5=-0.039$ for the first five excitonic states.
They correspond to the following series of the exciton binding energies $E_1=-235$\,meV, $E_2=-68$\,meV, $E_3=-31$\,meV, $E_4=-18$\,meV, $E_5=-11$\,meV.
Note that the binding energies of $s$ excitons in the bilayer are a little bit larger compared to their monolayer counterparts. 
This can be explained by the fact that the effective dielectric constant for the bilayer is smaller than that for the monolayer. 
The energy difference between 1$s$ and 2$s$ excitons is $|\Delta E_{12}|=167$\,meV, which is significantly larger than the experimental value $|\Delta E^{exp}_{12}|=118\pm6$\,meV. 
Therefore, the model with $\epsilon_\perp=1$ is not consistent with the experiment.

 \subsection{Bilayer: $\epsilon_\perp=\epsilon_\perp^{bil}$} 

In the previous section, we have demonstrated that the model with $\epsilon_\perp=1$ is not able to predict correctly the 
parameter $\Delta E_{12}$ for the BL S-TMD crystal. This deviation can be explained in two possible cases. 
The first case is based on the oversimplification of the model. In this model, we operate with infinite thin layers and distances between them. However, in the S-TMD crystal, the thickness of its constituents (monolayers) is comparable with the distance between the layers. One of the ways to keep the simplicity of the previous model but take into account
 the more complex electric response of few-layers in perpendicular to the layers' direction is to introduce the out-of-plane dielectric constant $\epsilon_\perp>1$. The second case is based on the observation that due to encapsulation the splitting $\Delta_\text{v}$ is significantly reduced \cite{Gong2013, Slobodeniuk2019}. The reduction of 
$\Delta_\text{v}$ leads to a more homogeneous distribution of the hole charge in the bilayer in the perpendicular 
direction, which reduces the Coulomb interaction between the electron and hole quasiparticles. 
We first calculate the binding energies for $\epsilon_\perp^{bil}=7.7$ \cite{Laturia2018}. 
If this calculation will not give the appropriate answer, we should conclude the renormalisation of the $\Delta_\text{v}$ parameter. 

In order to obtain the expression for the potential with $\epsilon_\perp=\epsilon_\perp^{bil}\neq 1$, we 
make the replacements $k\rightarrow k/\sqrt{\epsilon_\perp}$, $Q\rightarrow Q/\epsilon_\perp$, 
$\varepsilon\rightarrow \varepsilon/\sqrt{\epsilon_\perp}$, with $\epsilon_\perp=7.7$ \cite{Laturia2018} in 
Eqs.~(\ref{eq:phi_1}) and (\ref{eq:phi_2}).
This replacement effectively changes the parameters 
$L\rightarrow \widetilde{L}=L/\sqrt{\epsilon_\perp}$, 
$r_0\rightarrow \widetilde{r}_0=r_0/\sqrt{\epsilon_\perp}$,
$\delta\rightarrow \widetilde{\delta}=(\varepsilon-\sqrt{\epsilon_\perp})/(\varepsilon+\sqrt{\epsilon_\perp})$
in the expression for the full potential (\ref{eq:v_bil})
\begin{align}
v_{bil}(\xi)=2\varepsilon\int_0^\infty dx J_0(x\xi)
\frac{0.932(1-\widetilde{\delta})\left(e^{2xl}[(1-\widetilde{\delta})\varepsilon x+1]-[\varepsilon x(1-\widetilde{\delta})+\widetilde{\delta}]\right)
+0.068(1-\widetilde{\delta})^2e^{xl}}
{e^{2xl}[(1-\widetilde{\delta})\varepsilon x+1]^2-[\widetilde{\delta}+\varepsilon x(1-\widetilde{\delta})]^2}.
\end{align}
Note that the dimensionless parameter $l=\widetilde{L}\varepsilon/\widetilde{r}_0=L\varepsilon/r_0=0.56$ remains the same. 
The eigenvalue equation in dimensionless coordinate $\xi=\rho\varepsilon/\widetilde{r}_0$ becomes 
\begin{equation}
\Big\{b^2\epsilon_\perp \frac{1}{\xi}\frac{d}{d\xi}\Big(\xi\frac{d}{d\xi}\Big)+
bv_{bil}(\xi)+\epsilon\Big\}\psi(\xi)=0,
\end{equation} 
with the same notations for $\epsilon$ and $b$ as in the previous subsections.
The eigenvalues are $\epsilon_1=-0.599$, $\epsilon_2=-0.178$, $\epsilon_3=-0.089$, $\epsilon_4=-0.057$, and $\epsilon_5=-0.04$ for the first five excitonic states.
It gives us the energies of the ground (1$s$) and the four excited (2$s$, 3$s$, 4$s$, and 5$s$) states equal to $E_1=-177$\,meV, $E_2=-53$\,meV, $E_3=-26$\,meV, $E_4=-17$\,meV, $E_5=-12$\,meV, respectively.
The energy difference between 1$s$ and 2$s$ excitons is $|\Delta E_{12}|=124$\,meV, which is consistent with the experimental value $|\Delta E^{exp}_{12}|=118\pm6$\,meV.

We calculate numerically the wave functions in momentum space using the formula 
\begin{equation}
\psi_{ns}(k)=\int_0^\infty d\rho \rho J_0(k\rho)\psi_{ns}(\rho)=\Big(\frac{\widetilde{r}_0}{\varepsilon}\Big)^2
\int_0^\infty d\xi \xi J_0\Big(k\frac{\widetilde{r}_0}{\varepsilon}\xi\Big)\psi_{ns}(\xi),
\end{equation} 
where $\psi_{ns}(r)$ is the coordinate dependent wave-function for $ns$ state, and $\psi_{ns}(\xi)=\psi_{ns}(\rho\varepsilon/\widetilde{r}_0)$. 
The corresponding approximations for wave functions in momentum space are 
\begin{equation}
\psi_{1s}(k)/\psi_{1s}(0)\approx \frac{1}{\left(3.00818 (k\widetilde{r}_0/\varepsilon)^2+1\right)^{2.08207}},
\end{equation}
with $\psi_{1s}(0)=1.74062 \widetilde{r}_0/\varepsilon$, and 
\begin{equation}
\psi_{2s}(k)/\psi_{2s}(0)\approx \frac{1-22.4166 (k\widetilde{r}_0/\varepsilon)^2}
{(10.1202 (k\widetilde{r}_0/\varepsilon)^2+1)^{3.24956}},
\end{equation} 
with $\psi_{2s}(0)=5.26663 \widetilde{r}_0/\varepsilon$.
Using the aforementioned we can estimate the size of the wave functions in momentum space as 
\begin{equation}
k_{1s}\approx \frac{0.576565}{\widetilde{r}_0/\varepsilon}, \quad k_{2s}\approx \frac{0.314344}{\widetilde{r}_0/\varepsilon},
\end{equation}
where $\widetilde{r}_0/\varepsilon \approx 4.14\mbox{\AA}$. 

To fulfill the study, we also estimate the binding energies and wave functions of the ground state of interlayer A (IL), 
intralayer B (1s$^\text{B}$), and interlayer B excitons.   

The binding energy and wave function for the interlayer A exciton (IL) in momentum space are 
$E_1=-118$\,meV, and  
\begin{equation}
\psi_{1s}(k)/\psi_{1s}(0)\approx \frac{1}{\left(3.70421 (k\widetilde{r}_0/\varepsilon)^2+1\right)^{3.09065}},
\end{equation}
with $\psi_{1s}(0)=2.47168 \widetilde{r}_0/\varepsilon$.
The size of the corresponding wave function in momentum space is
\begin{equation}
k_{1s}\approx \frac{0.51958}{\widetilde{r}_0/\varepsilon}, 
\end{equation}
where $\widetilde{r}_0/\varepsilon \approx 4.14\mbox{\AA}$. 

The binding energy and wave function for the intralayer B exciton (1s$^\text{B}$) in momentum space are
$E_1=-190$\,meV, and  
\begin{equation}
\psi_{1s}(k)/\psi_{1s}(0)\approx \frac{1}{\left(2.41494(k\widetilde{r}_0/\varepsilon)^2+1\right)^{2.11039}},
\end{equation}
with $\psi_{1s}(0)= 1.57347\widetilde{r}_0/\varepsilon$.
The size of the corresponding wave function in momentum space is
\begin{equation}
k_{1s}\approx \frac{0.643497}{\widetilde{r}_0/\varepsilon}.
\end{equation}

The binding energy and wave function for the interlayer B exciton in momentum space are
$E_1=-107$\,meV, and  
\begin{equation}
\psi_{1s}(k)/\psi_{1s}(0)\approx \frac{1}{\left(4.47643(k\widetilde{r}_0/\varepsilon)^2+1\right)^{3.00066}},
\end{equation}
with $\psi_{1s}(0)= 2.66952\widetilde{r}_0/\varepsilon$.
The size of the corresponding wave function in momentum space is
\begin{equation}
k_{1s}\approx \frac{0.472644}{\widetilde{r}_0/\varepsilon}.
\end{equation}

\section{Exciton $g$-factors \label{sec:results}}
Table~\ref{tab:g-si} presents the calculated values of the band $g$-factors for higher-energy c+1 (v) and lower-energy c (v-1) subbands in CB (VB) at the K$^+$ point. 
The values at the K$^-$ point are opposite due to time-reversal symmetry.  
These band $g$-factors, inserted in Eq. 15 in the main text, yield the $g$-factors of band-to-band transitions at the $\mathbf{k}$=K$^\pm$ point. 
Due to the $k$-dependence of band $g$-factors and exciton wavefunction spreads (see Fig. 4 in the main text), the $g$-factors of excitonic states are smaller in magnitude. 
However, values from Tab.~\ref{tab:g-si} can serve as an initial estimation of $g$-factors of different excitons formed in van der Waals heterostructures comprising the ML and BL MoSe$_2$.

\begin{table}[h!]
\caption{\label{tab:g-si}
Theoretical $g$-factors of valence and conduction states at K$^+$ point in ML and BL MoSe$_2$.}
\begin{tabular}{ccccc}
      & $g_{v-1}$ & $g_v$  & $g_c$ & $g_{c+1}$ \\
\hline
ML & 2.31 & 4.46 & 2.48 & 0.30 \\
BL & 2.22 & 4.35 & 2.36 & 0.24 \\
\hline
\end{tabular}
\end{table}